\documentclass[ss,preprint]{imsart}
\RequirePackage[utf8]{inputenc}
\RequirePackage[T1]{fontenc}
\RequirePackage{lmodern}
\RequirePackage{amsmath,amsthm,amsfonts,amssymb,bm}
\RequirePackage{dsfont}			
\RequirePackage{enumitem}
\RequirePackage{graphicx}
\RequirePackage{booktabs}
\RequirePackage{upgreek}
\RequirePackage{multirow}
\RequirePackage{makecell}
\RequirePackage{colortbl}
\RequirePackage{subfig}
\RequirePackage{url}
\RequirePackage[font={small,it}]{caption}
\RequirePackage{natbib}
\RequirePackage[colorlinks,citecolor=blue,urlcolor=blue]{hyperref}

\usepackage[scaled=0.8]{beramono}

\doi{https://doi.org/10.1214/18-SS119}
\pubyear{2018}
\volume{12}
\firstpage{1}
\lastpage{48}

\startlocaldefs
\newcommand{\MU}{\bm{\mu}}
\newcommand{\SIG}{\bm{\Sigma}}
\newcommand{\X}{\mathbf{X}}
\newcommand{\Xb}{\mathbb{X}}
\newcommand{\x}{\mathbf{x}}
\newcommand{\z}{\mathbf{z}}
\newcommand{\Q}{Q_{\lambda}(\bm{\Theta})}
\newcommand{\sumj}{\sum_{j=1}^J}
\newcommand{\sumg}{\sum_{g=1}^G}
\newcommand{\sumi}{\sum_{i=1}^N}
\newcommand{\vbar}{\,\lvert\,}
\newcommand{\vphi}{\bm{\varphi}}

\endlocaldefs

\begin{document}

\begin{frontmatter}

\title{Variable Selection Methods for Model-based Clustering}
\thankstext{t1}{This work was supported by the Science Foundation Ireland funded Insight Research Centre (SFI/12/RC/2289)}
\runtitle{Variable Selection Methods}

\author{\fnms{Michael} \snm{Fop}\thanksref{t1}\corref{}\ead[label=e1]{michael.fop@ucd.ie}}
\and
\author{\fnms{Thomas Brendan} \snm{Murphy}\thanksref{t1}\ead[label=e2]{brendan.murphy@ucd.ie}}

\address{University College Dublin\\
\printead{e1,e2}}

\runauthor{Fop, M. and Murphy, T. B.}

\begin{abstract}
%
Model-based clustering is a popular approach for clustering multivariate data which has seen applications in numerous fields. Nowadays, high-dimensional data are more and more common and the model-based clustering approach has adapted to deal with the increasing dimensionality. In particular, the development of variable selection techniques has received a lot of attention and research effort in recent years. Even for small size problems, variable selection has been advocated to facilitate the interpretation of the clustering results. This review provides a summary of the methods developed for variable selection in model-based clustering. Existing R packages implementing the different methods are indicated and illustrated in application to two data analysis examples.
\end{abstract}




\tableofcontents

\end{frontmatter}

\section{Introduction}
\label{intro}
Model-based clustering is a well established and popular tool for clustering multivariate data. In this approach, clustering is formulated in a modeling framework and the data generating process is represented through a finite mixture of probability distributions. Often, all the variables at a user's disposal are employed in the modeling. Nonetheless, in many situations considering all the variables unnecessarily increases the model complexity. Moreover, some variables may not possess any clustering information and are of no use in the detection of the group structure. Rather, they could be detrimental to the clustering. Likewise, the case were all of the variables contain clustering information can also be problematic. Along with the increasing number of dimensions comes the \emph{curse of dimensionality} \citep{bellman:1957} and including superfluous variables in the model leads to identifiability problems and over-parameterization \citep{bouveyron:2014,bartholomew:knott:2011}. Therefore, resorting to variable selection techniques can facilitate model fitting, ease the interpretation of the results and lead to data classification of better quality. Even in situations of moderate or low dimensionality, reducing the set of variables employed in the clustering process can be beneficial \citep{fowlkes:1988}. 

This article gives a review of the available methods for variable selection in model-based clustering. Starting from early works on the topic, we will give a summary of the various approaches up to the most recent developments. The focus will be on variable selection when clustering multivariate continuous and categorical data, the two most common data types in practice. References to the existing \textsf{R} packages that implement the different methods are provided. As illustrative examples, we will apply the variable selection methods on two illustrative datasets.

The exposition is structured as follows. Section~\ref{clustervarsel} introduces the main features of the variable selection methods for mixture model clustering. Section~\ref{mbc} recalls the model-based clustering framework, with a brief description of the two principal models for multivariate continuous and categorical data: Gaussian mixture models and latent class analysis. In Section~\ref{vsgmm}, classical and recent variable selection methods for Gaussian mixture models are reviewed. The section terminates with the list of available \textsf{R} packages and a data analysis example concerning historical mortality rates. Section~\ref{vslca} is dedicated to variable selection methods for latent class analysis. \textsf{R} packages and an application to voting data are presented in conclusion. The paper ends with a discussion in Section~\ref{disc}.

\section{Methods of variable selection}
\label{clustervarsel} 
When performing variable selection for clustering, the goal is to retain only the relevant variables. With this aim, it is crucial to define what it means for a variable to be, or not to be, ``relevant''. In supervised learning the matter has been the object of rigorous discussion for a long time; we mention the works of \cite{yu:2004,blum:1997,kohavi:1997,koller:1996} and \cite{john:1994}. Within the model-based clustering approach, the question has been addressed recently in \cite{ritter:2014}. Model-based clustering places the clustering task into a formal modeling framework and the group structure is embedded in the group membership variable \citep{mclachlan:1988}. In this case, the definition of relevance can be expressed in terms of probabilistic dependence (or independence) statements with respect to this variable \citep{ritter:2014}. \emph{Relevant variables} contain the essential clustering information. In a model-based clustering context, their distribution directly depends on the group membership variable. On the other hand, \emph{irrelevant variables} do not convey any beneficial information. These can be further divided into redundant and uninformative variables. \emph{Redundant variables} provide information of similar quality to that already available in the relevant ones, therefore are not needed for a parsimonious modeling. In many situations, they contain similar information because they are correlated with the relevant ones. In terms of distributional representation, we may think of redundant variables as conditionally independent of the grouping variable given the relevant variables; they could be useful for clustering, but only if the relevant ones are not present. On the contrary, \emph{uninformative variables} possess no discriminative information whatsoever. They correspond to noise and their distribution is completely independent of the group structure. 

Different model-based clustering and variable selection strategies can be delineated by distributional assumptions on relevant and irrelevant variables. Two main assumptions are peculiar to the task of variable selection and mixture model clustering; these are:
\begin{itemize}
 \item {\em Local independence assumption}. The relevant variables are conditionally independent within the groups. 
 \item {\em Global independence assumption}. The irrelevant variables are independent of relevant clustering variables. 
\end{itemize}
\begin{figure}[!t]
\centering
    \subfloat[][\em Local independence assumption]
    {\includegraphics[scale=1.4]{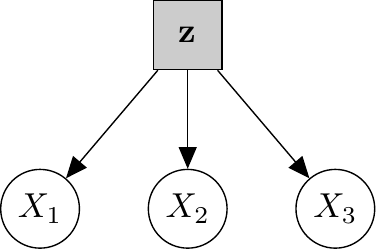}} \qquad\qquad
    \subfloat[][\em Global independence assumption]
    {\includegraphics[scale=1.4]{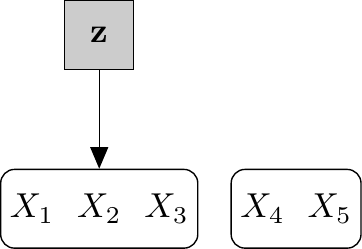}} \\
    \caption{\label{graph0} Local and global independence assumptions. In the example, $\z$ is the group membership variable, $X_1,X_2$ and $X_3$ are relevant clustering variables, while $X_4$ and $X_5$ are irrelevant and not related to $\z$. Under the local independence assumption there are no edges among the relevant variables. Under the global independence assumption there is no edge between the set of relevant variables and the set of irrelevant ones.}
\end{figure}
Figure~\ref{graph0} gives a simple sketch of the two independence statements. In specifying a general strategy for model-based clustering and selection of variables, these assumptions can be combined together or employed separately. They have different implications on the clustering model and the model for the relations among the variables. In particular, the local independence assumption helps to simplify the modeling of the joint distribution of relevant variables and is especially useful in high-dimensional data settings. It is a standard assumption of the latent class analysis model \citep{clogg:1988}. For Gaussian mixture models, the statement corresponds to assuming components with diagonal covariance matrices. The global independence assumption implies that the joint distribution of the variables factors into the product of the mixture distribution of the relevant variables and the distribution of the irrelevant variables. The term ``global'' is used because the independence statement affects the distribution of \emph{all} the variables, not only the clustering ones. This assumption simplifies the modeling of the relation between relevant and irrelevant variables. However, it limits the capability of taking into account the presence of redundant variables \citep{law:figueiredo:2004,raftery:dean:2006}. Many of the methods that are presented in this review make use of one of the two assumptions above, either implicitly or explicitly.

Likewise, how the variable selection algorithm interacts with the model fitting process defines the overall approach to the problem. For a general learning task, the principal distinction is in whether the selection is carried out separately or jointly to the learning procedure \citep{john:1994,dash:1997,dy:brodley:2004,sayes:2007,liu:2007}. The first case corresponds to \emph{filter methods}, where the selection is performed as a pre (or sometimes post) processing step. The second corresponds to \emph{wrapper methods}, that combine learning and variable selection at the same time. In this case the selection procedure is ``wrapped'' around the learning algorithm. Filter approaches are easy to implement and computationally efficient. However, wrapper methods often provide superior results, despite being more involved \citep{blum:1997,kohavi:1997,guyon:2003}. 

In model-based clustering, filter methods perform the variable selection before (or after) the model has been estimated. The inferred classification is then used to evaluate the quality of the variables. In contrast, with a wrapper method model estimation and selection are conducted simultaneously. Variables selected via filter methods can miss important information as the selection is external to the model estimation. Indeed, wrapper methods have recently attracted most of the attention and are the most widespread. Their advantages lie in the fact that they can be naturally included in the model fitting, can lead to a better classification and can provide a better representation of the data generating process \citep{dy:brodley:2004,law:figueiredo:2004}. 

Within the class of wrappers, the various methods for variable selection in mixture model clustering can be further distinguished according to the type of statistical approach used. Three major approaches can be found in the literature:
\begin{itemize}
 \item {\em Bayesian approaches}. In this class of methods, the problem of variable selection is addressed assuming the existence of a latent variable indicating if an observed variable characterizes a mixture distribution or not. Variable selection is conducted by making inference about the posterior distribution of such a latent variable. 
 \item {\em Penalization approaches}. Within this category, variable selection is performed by using a penalized log-likelihood approach. The penalization term is a function of the mixture parameters and acts to shrink the estimates towards an overall common value. Variables whose estimates take this common value across the different mixture components are considered irrelevant and are discarded.
 \item {\em Model selection approaches}. Here the task of variable selection is re-formulated as a model selection problem. Different models are specified according to the role of the variables towards the clustering structure. Consequently, relevant variables are selected by comparing different models using some predefined criterion. 
\end{itemize}
The above categorization is not exhaustive nor the definitions are intended to be mutually exclusive. Indeed, many of the methods that will be presented in the rest of the paper have some degree of overlap and a method belonging to one type of approach could be easily rephrased in terms of the other ones. These three general approaches are the predominant in the literature and the classification will be used for ease of exposition and a systematic presentation. 

In summary, the existing variable selection methods for model-based clustering are differentiated in relation to the distributional assumptions for relevant and irrelevant variables, the interaction between variable selection and model fitting, and the general statistical approach. We will give an overview of these methods after a short description of the model-based clustering framework.

\section{Model-based clustering}
\label{mbc}
Let $\X$ be the $N \times J$ data matrix, where each row $\x_i = (x_{i1},\, \dots,\, x_{ij},\, \dots,\, x_{iJ})$ is the realization of a $J$-dimensional vector of random variables $\Xb = (X_1,\,\dots,\,X_j,$ $\dots, \, X_J)$. Model-based clustering assumes that each observation arises from a finite mixture of $G$ probability distributions, each representing a different cluster or group \citep{fraley:raftery:2002,melnykov:2010,mcnicholas:2016}. The general form of a finite mixture distribution is specified as follows:
\begin{equation}\label{eq:1}
p(\x_i; \bm{\Theta}) = \sumg \tau_g \, p(\x_i ; \bm{\Theta}_g),
\end{equation} 
where the $\tau_g$ are the mixing probabilities and $\bm{\Theta}_g$ is the parameter set corresponding to component $g$; $\bm{\Theta}$ denotes the set of all parameters of the mixture.  The component densities fully characterize the group structure of the data and each observation belongs to the corresponding cluster according to a latent cluster membership indicator variable $\z_i = (z_{i1},\,\dots, \, z_{ig},\,\dots,z_{iG})$, such that $z_{ig} = 1$ if $\x_i$ arises from the $g$th subpopulation \citep{mclachlan:peel:2000,mclachlan:1988}.

For a fixed number of components, parameters are usually estimated using the EM algorithm \citep{dempster:etal:1977,mclachlan:krishnan:2008,bartholomew:knott:2011,ohagan:2012}. Moreover, generally model selection corresponds to the selection of the number of components $G$ and to accomplish the task a plethora of methods have been suggested in the literature, the \emph{Bayesian Information Criterion} \citep[BIC,][]{schwarz:1978} being the most popular one. Another popular approach for mixture model selection is the \emph{Integrated Complete-data Likelihood} criterion \citep[ICL,][]{biernacki:2000}, which gives more importance to models with well separated clusters. See \cite{mclachlan:rathnayake:2014} for a detailed review of the various methods. 

After parameters have been estimated, each observation is assigned to the corresponding cluster using the \emph{maximum a posteriori} (MAP) rule \citep{mclachlan:peel:2000,mcnicholas:2016}. The posterior probabilities $u_{ig} = \Pr(z_{ig} = 1 \vbar \x_i)$ of observing cluster $g$ given the data point $i$ are estimated as follows:
$$
\hat{u}_{ig} = \dfrac{\hat{\tau}_g \, p(\x_i ; \hat{\bm{\Theta}}_g)}{\sum_{h=1}^G \hat{\tau}_h \, p(\x_i ; \hat{\bm{\Theta}}_h)},
$$
Then observation $\x_i$ is assigned to cluster $g$ if 
$$
g = \text{MAP}(\hat{\textbf{u}}_{i}) =  \underset{h}{\operatorname{argmax}} ~ \lbrace \hat{u}_{i1},\, \dots,\, \hat{u}_{ih},\, \dots,\, \hat{u}_{iG} \rbrace.
$$

According to the nature of the data, different specifications for the component densities in~\eqref{eq:1} have been proposed. In the following sections we focus on the cases of continuous and categorical data, taking in consideration the two most popular distributions: Gaussian and Multinomial. However, various and more flexible distributional assumptions can be specified, enabling to take into account for skewness, heavy tails and different data types; for example, see \cite{mcnicholas:2016} for a review on non-Gaussian components, \cite{mclachlan:1998} for mixtures of multivariate \emph{t}-Student distributions, \cite{lee:2013} and \cite{lee:2016} for mixtures of skew-$t$ and skew normal distributions, \cite{karlis:2007} and \cite{rau:2015} for clustering multivariate count data, \cite{mcparland:2016} for mixed data model-based clustering, \cite{Kosmidis:2016} for the use of copulas in model-based clustering, \cite{desantis:2008} for latent class analysis of ordinal data. 

\subsection{Gaussian mixture model}
When clustering multivariate continuous data, a common approach is to model each component density by a multivariate Gaussian distribution. For a {\em Gaussian mixture model} (GMM) the mixture density in~\eqref{eq:1} becomes:
\begin{equation}\label{gmm}
p(\x_i; \bm{\Theta}) = \sumg \tau_g \, \phi(\x_i ; \MU_g, \SIG_g),
\end{equation}
where $\phi$ is the multivariate Gaussian density and $\MU_g$ and $\SIG_g$ are the mean and covariance parameters respectively; see \cite{fraley:raftery:2002}, \cite{melnykov:2010} and \cite{mcnicholas:2016} for reviews. To attain parsimony, several approaches involving re-parameterizations of the covariance matrix $\SIG_g$ have been presented; for example \cite{banfield:raftery:1993,celeux:govaert:1995,bouveyron:2007,mcnicholas:murphy:2008,biernacki:lourme:2014}. We refer to \cite{bouveyron:2014} for a review. 

Model based-clustering via GMMs can be performed in \textsf{R} \citep{R} using the packages \texttt{mclust} \citep{scrucca:etal:2016}, \texttt{Rmixmod} \citep{lebret:2015}, \texttt{EMCluster} \citep{chen:2015}, \texttt{mixtools} \citep{mixtools} and \texttt{flexmix} \citep{leisch:2004}; also the \texttt{EMMIX} software \citep{mclachlan:99} can be used for fitting mixtures of Gaussian distributions. Lastly, we note that in \textsf{Python} \citep{python}, GMMs estimation can be executed via the packages \texttt{scikit-learn}\footnote{\url{http://scikit-learn.org/stable/}} and \texttt{PyMixmod} \footnote{\url{http://www.mixmod.org/spip.php?article62}}.

\subsection{Latent class analysis model}
For clustering multivariate categorical data, the common approach is to use the {\em latent class analysis model} \citep[LCA,][]{bartholomew:knott:2011}. Under this model, the mixture density in~\eqref{eq:1} is a mixture of Multinomial distributions as follows:
\begin{equation}\label{lca}
p(\x_i; \bm{\Theta}) = \sumg \tau_g \, \mathcal{C}(\x_i ; \bm{\pi}_g),
\end{equation}
where 
$$
\mathcal{C}(\x_i ; \bm{\pi}_g) = \prod_{j=1}^J \prod_{c=1}^{C_j} \pi_{gjc}^{\mathds{1}\lbrace x_{ij} = c \rbrace},
$$
with $\pi_{gjc}$ representing the probability of occurrence of category $c$ for variable $X_j$ in class $g$, and $C_j$ the number of categories of variable $j$. The factorization in $\mathcal{C}(\x_i ; \bm{\pi}_g)$ is due to the local independence assumption, stating that the variables are independent within each latent class \citep{clogg:1988}. More details about the model can be found in \cite{vermunt:magdison:2002,agresti:2002,collins:2010}. For different values of $G$, not all the models can be fitted and constraints on the parameters need to be placed in order to ensure identifiability \citep{goodman:1974}; for examples see \cite{formann:1985}.

\textsf{R} packages implementing the LCA model for clustering categorical data are \texttt{BayesLCA} \citep{white:2014}, \texttt{poLCA} \citep{linzer:2011} and \texttt{flexmix} \citep{leisch:2004}; package \texttt{e1071} \citep{e1071} contains function \texttt{lca} for fitting the LCA model on binary data.

\section{Variable selection methods for Gaussian mixture models}
\label{vsgmm}
%

In this section we provide an overview of the available methods for clustering and variable selection using Gaussian mixture models. \cite{steinley:2008} and \cite{celeux:martin:2014} compare and evaluate the performances of some of the methods described in the subsequent sections. In particular, \cite{steinley:2008} perform an empirical comparison of different procedures for clustering and variable selection, also those not based on mixture models. While \cite{celeux:martin:2014} compare the model selection approach of \cite{maugis:celeux:2009:a} (see Section \ref{vsgmm_modsel}) and the regularization approach of \cite{witten:tibshirani:2010} (see Section \ref{vsgmm_pen}); the authors found that, in the case of correlated variables, the model selection approach was substantially more accurate in terms of both classification and variable selection than the regularization approach, and that both gave more accurate classifications than K-means without variable selection.

\subsection{Bayesian approaches}
\label{vsgmm_bayes}
Various methods have been developed within the Bayesian paradigm for simultaneous model-based clustering and variable selection. A common feature among them is the introduction of a variable $\vphi$, usually following a distribution $p(\vphi)$. The variable splits $\X$ into two sets: $\X^C$, the set of variables defining a Gaussian mixture distribution, and $\X^{NC}$, the set of variables indicating a single multivariate Gaussian distribution. In its most general form, the distribution of the data, conditional on $\z$, can be expressed as:	
$$
p(\X \vbar \z, G, \vphi, \bm{\Omega}) = p(\X^C \vbar \z, G, \vphi, \bm{\Theta}) \, p(\X^{NC} \vbar \vphi, \bm{\Gamma}),
$$
where $p(\X^C \vbar \z, G, \vphi, \bm{\Theta})$ is a mixture distribution whose parameters are denoted by $\bm{\Theta}$, $p(\X^{NC} \vbar \vphi, \bm{\Gamma})$ is a single multivariate Gaussian distribution with parameters $\bm{\Gamma}$, and $\bm{\Omega}$ denotes the collection $\lbrace \bm{\Theta}, \bm{\Gamma} \rbrace$ of all parameters. Then, with the aim of variable selection and clustering, the focus is in drawing inference from the posterior distribution:
$$
p(\z, G, \vphi, \bm{\Omega} \vbar \X) \propto p(\X \vbar \z, G, \vphi, \bm{\Omega}) \, p(\z \vbar \bm{\Omega}) \, p(G) \, p(\vphi) \, p(\bm{\Omega}).
$$

In this context, \cite{liu:2003} propose the $\emph{anchor mode model}$, where variable selection is performed by selecting the most informative principal components (or factors) of the data. In this approach, a preliminary dimension reduction through principal component analysis is performed and the first $k_0$ factors are retained. Then it is assumed that only a subset of these factors is informative for clustering and distributed according to a mixture of Gaussians, while the remaining components follow a simple Gaussian distribution. The subset consists of the first $\varphi$ principal components, where $\varphi$ is a random variable distributed according to the prior distribution $p(\varphi)$. Inference on this number of relevant factors is conducted employing a \emph{Markov Chain Monte-Carlo} (MCMC) scheme where the prior $p(\varphi)$ is taken to be the Uniform distribution. The method is shown to perform well on high-dimensional gene expression data, however, selection is performed on a set of features derived from the original variables and in general the principal components with the larger eigenvalues will not necessarily contain the most useful information about the clustering \citep[see][]{chang:1983}.

As an alternative to the ``hard selection'' approach (a variable is either selected or not), \cite{law:figueiredo:2004} suggest a Bayesian framework where the concept of \emph{feature saliency} is introduced. Let $\vphi = (\varphi_1,\, \dots,\, \varphi_j,\, \dots,\, \varphi_J)$ be a binary variable such that $\varphi_j = 1$ if $X_j$ is relevant and $\varphi_j=0$ otherwise. Then the saliency of variable $X_j$ is the quantity $\rho_j = \Pr(\varphi_j = 1)$ and can be interpreted as the importance of the variable in characterizing the cluster structure of the data. Assuming conditional independence of the variables, it follows that the likelihood of a data-point can be expressed as:
$$
p(\x_i \vbar \bm{\Omega}) = \sumg \tau_g \prod_{j=1}^J \left[\, \rho_j\, \phi(x_{ij} \vbar \mu_g, \sigma^2_g) \,+\, (1-\rho_j)\, \phi(x_{ij} \vbar \bm{\Gamma}) \,\right],
$$
with clear use of the notation. Rather than $\vphi$, the interest here is in recovering the probabilities $\rho_j$ and a Dirichlet prior is placed on the corresponding vector. To encourage the saliences of some variables to converge to zero, the authors adopt a \emph{minimum message length} criterion \citep{wallace:1987} and utilize an EM algorithm for maximum a posteriori estimation. Within the same framework, \cite{const:2006} consider a variational learning method for estimation of the saliences. 

The same idea of a binary clustering-relevance indicator variable $\vphi$ is employed in \cite{tadesse:sha:2005}. The authors assume a prior on $\vphi$ of the form:
$$
p(\vphi \vbar \eta) = \prod_{j=1}^J \eta^{\varphi_j} (1-\eta)^{1-\varphi_j},
$$
with $\eta$ the hyper-parameter interpreted as the proportion of variables expected to discriminate the groups. A MCMC scheme is used for inference, and the vector $\vphi$ is updated using a Metropolis search where a new candidate is generated from the previous state by adding, removing and swapping at random its entries. Posterior inference on $\vphi$ is drawn after integrating out the parameters and considering the marginal posterior $p(\varphi_j = 1 \vbar \X)$. Then the best clustering variables can be
identified as those with largest marginal posterior, $p(\varphi_j = 1 \vbar \X) > t$, with a specified $t$. Alternatively, the selection can be performed by taking into account the complete vector $\vphi$ with largest posterior probability among all visited vectors throughout the chain, thus considering the joint density of $\vphi$. The method is shown to perform well in clustering high-dimensional microarray data. Furthermore, in subsequent work, \cite{kim:tadesse:2006} extend the approach by formulating the clustering in terms of an infinite mixture of Gaussian distributions via Dirichlet process mixtures, while \cite{swartz:2008} expand it to the modeling of data with a known structure imposed from an experimental design.

\subsection{Penalization approaches}
\label{vsgmm_pen}
In this context, a penalization term is introduced on the model parameters and variable selection is performed by inducing sparsity in the estimates. The aim is to maximize a penalized version of the log-likelihood under a Gaussian mixture model and discard those variables whose parameter estimates are shrunken to zero or to a common value across the mixture components. In its general form, this penalized log-likelihood is as follows: 
\begin{equation}\label{eq:2}
 \ell_Q = \sumi \log \left\lbrace \sumg \tau_g \, \phi(\x_i ; \bm{\Theta}_g) \right\rbrace - \Q,
\end{equation}
where the penalization term $Q_{\bm{\lambda}}(\bm{\Theta})$ is a function of the Gaussian densities parameters $\bm{\Theta}$ and $\lambda$, a generic penalty parameter (here in the notation we denoted with $\bm{\Theta}$ the collection of all Gaussian density parameters and with $\bm{\Theta}_g$ the subset corresponding to component $g$). 
Generally, the various methods are differentiated by the form of the function $Q_{\bm{\lambda}}(\cdot)$, having different implications on the selection of variables. 

Seminal work in this class of approaches is the method introduced by \cite{pan:shen:2007}. The authors use a $L_1$ penalty function of the form:
$$
\lambda \sumg \sumj \,\, \vbar \mu_{gj} \vbar.
$$
After centering the data, the method realizes variable selection by automatically shrinking several small estimates of $\mu_{gj}$ towards zero. Indeed, if $\mu_{gj} = 0$ for all $g$, the component means for variable $j$ are equal to the overall data mean and variable $j$ does not contribute to the clustering. The authors show an application to the \cite{golub} leukemia gene expression dataset, demonstrating the usefulness of the approach in ``high dimensions -- small sample size'' settings.

A closely related approach is the one suggested by \cite{bhat:2014}, where the penalizing function accounts for the size of the clusters and is given by:
$$
\lambda \sumg N \tau_g \sumj \,\, \vbar \mu_{gj} \vbar.
$$
The parameter $\lambda$ depends on the sample size and the authors derive a BIC-type model selection criterion for high-dimensional settings.

The $L_1$ penalty function treats each $\mu_{gj}$ individually, not using the information that, across the mixture components, the parameters $(\mu_{1j},\, \dots,\, \mu_{gj},\, \dots,\, \mu_{Gj})$ corresponding to the same variable $X_j$ are grouped. This results in the fact that, if for a fixed variable $j$ and some component $g$, we have $\mu_{gj} \neq 0$ while $\mu_{kj} = 0$ for all the remaining components, then the variable would not be excluded. \cite{wang:zhu:2008} suggest a solution to the problem by replacing the $L_1$ norm with the $L_{\infty}$ norm. Thus the penalty function is given by:
\begin{equation*}
\lambda \sumj \underset{g}{\operatorname{max}}\, \lbrace \vbar\mu_{1j}\vbar,\, \dots,\, \vbar\mu_{gj}\vbar,\, \dots,\, \vbar\mu_{Gj}\vbar \rbrace.
\end{equation*}
After re-parameterizing $\mu_{gj} = \alpha_j \beta_{gj}\,\, (\alpha_j \geq 0)$, the authors consider also a hierarchical penalization function of the form:
\begin{equation*}
\lambda_1 \sumj \alpha_j + \lambda_2 \sumg \sumj \,\, \vbar \beta_{gj} \vbar,
\end{equation*}
with $\lambda_1$ controlling the amount of shrinkage on the $\mu_{gj}$ as a group for $g =1,\,\dots,G$, and $\lambda_2$ controlling the shrinkage effect within variable $j$. This function has the advantage of being more flexible and inducing a less ``hard'' penalization than the $L_{\infty}$ norm. Both penalty functions take into account the fact that component means corresponding to the same variable can be treated as grouped and tend to conduct a more effective variable selection.

The idea of grouped parameters is also accounted in \cite{xie:2008}. Here the authors suggest the use of two planes of grouping: \emph{vertical} and \emph{horizontal mean grouping}. For the first, mean parameters afferent to the same variable are treated as a whole and the penalty function is:
$$
\lambda \sqrt{G} \sumj\, \lvert\lvert\, \MU_j \,\lvert\lvert_2
$$
where $\MU_j = (\mu_{1j},\, \dots,\, \mu_{gj},\, \dots,\, \mu_{Gj})$ and $\lvert\lvert \cdot \lvert\lvert_2$ denotes the $L_2$ norm. The same idea is exploited in the horizontal grouping, where prior knowledge that some variables work in groups is introduced in the penalization. Here the variables can be grouped in $M$ groups indexed by $m$ and each one of size $H_m$. Then the function is given by:
$$
\lambda \sumg \sum_{m=1}^M \sqrt{H_m} \, \lvert\lvert\, \MU_{gm} \,\lvert\lvert_2,
$$
with $\MU_{gm}$ the vector of means of component $g$ for variables in group $m$. The two planes of penalization can be combined together and \cite{xie:2008} show their superior performance in comparison to the standard $L_1$ penalty.

An approach allowing to identify which variables are discriminative for which specific pairs of clusters is the one proposed by \cite{guo:2010}. They introduce a \emph{pairwise fusion} penalty of the form:
$$
\lambda \sumj \left( \, \sumg \sum_{h<g}\, \vbar \mu_{gj} - \mu_{gh} \vbar \, \right).
$$
The penalty function shrinks the mean parameters toward each other and if $\mu_{gj} = \mu_{gh}$, variable $X_j$ is not useful in discriminating components $g$ and $h$, but may be useful in discriminating the other components. A variable will be considered as non-informative and discarded only in the case where all the cluster means are shrunken to the same value.

A shared characteristic of the penalization methods presented to this point is the assumption that the data are generated by a Gaussian mixture distribution with a common diagonal covariance matrix. Consequently, the mixture components are differentiated only by their mean parameters and variables with cluster-specific means all equal can be discarded since non-informative to clustering. Nonetheless, in some situations, the assumption of a common isotropic covariance matrix might be too restrictive, as it implies that all the clusters are of spherical shape and have the same volume. For example, the \cite{golub} data contain a sample of 38 patients known to have three types of leukemia, with different variability in the gene expressions. In this case, assuming a common isotropic covariance matrix would likely lead to the selection of a number of clusters larger than the actual number of types of leukemia in order to accommodate for the extra within-cluster variability.

\cite{xie:pan:2008} move away from this assumption and present an approach for Gaussian mixture distributions with cluster-specific diagonal covariance matrices. They introduce two penalization terms of the form:
$$
\lambda_1 \sumg \sumj \,\, \vbar \mu_{gj} \vbar \,\, + \,\, \lambda_2 \sumg \sumj\,\, \vbar \sigma^2_{gj} - 1 \vbar.
$$
$$
\lambda_1 \sumg \sumj \,\, \vbar \mu_{gj} \vbar \,\, + \,\, \lambda_2 \sumg \sumj\,\, \vbar \log\sigma^2_{gj} \vbar.
$$
After the data have been standardized to have mean 0 and variance 1 for all the variables, variable selection is performed by removing those variables having common mean 0 and common variance 1 across the clusters. The authors also expand the penalization to account for grouped variables in a similar fashion as in \cite{xie:2008}. The method is applied to the \cite{golub} dataset and it is shown to perform a more parsimonious selection than the approach of \cite{pan:shen:2007} and to detect clusters where gene expressions have different variances.

\cite{zhou:2009} further extend the previous method to the case of unconstrained covariance matrices. The authors suggest the following penalization:
$$
\lambda_1 \sumg \sumj \,\, \vbar \mu_{gj} \vbar \,\, + \lambda_2 \sumg\,\, \lvert\lvert \mathbf{W}_{g} \,\lvert\lvert_1,
$$
where $\mathbf{W}_g$ is the precision matrix, $\mathbf{W}_g = \SIG_g^{-1}$, and $\lvert\lvert \mathbf{W} \lvert\lvert_1 = \sumj\sum_{h=1}^J \vbar w_{jh} \vbar$. The term involving the mean parameters is used for variable selection, while that for the precision is a regularization for dealing with high-dimensional data without imposing a diagonal covariance matrix. In this framework, covariances between the variables are taken into account and clusters are also allowed to be elliptical and of different shapes. The authors apply their method on the \cite{golub} data, improving the classification performance with respect to the previous approaches.

By introducing a penalization term in the mixture of factor analyzers model \cite[see][for example]{bouveyron:2014} dimension reduction and variable selection can be performed simultaneously. \cite{xie:2010} introduce a penalization approach in the mixture of factor analyzers model where the penalty function is given by:
$$
\lambda_1 \sumg \sumj \,\, \vbar \mu_{gj} \vbar \,+\, \lambda_2 \sumg \sumj \sqrt{ \sum_{r=1}^R\, \gamma_{gjr}^2 },
$$
with $\gamma_{gjr}$ the cluster-specific loading corresponding to variable $j$ and latent dimension $r$. The penalization on the loadings serves as a grouped variable penalty similarly to \cite{xie:2008}. The author apply the method for clustering gene expression data of lung cancer patients, uncovering clusters of subjects with distinct risks of mortality.

\cite{galimberti:2009} introduce a mixture of factor analyzers model with common factor loading matrix that projects the cluster means in a low dimensional space. They consider a penalized log-likelihood of the form:
$$
\ell_Q = \sumi \log \left\lbrace \sumg \tau_g \, \phi(\x_i ; \mathbf{V}\MU_g, \mathbf{V}\SIG_g\mathbf{V}' \,+\, \mathbf{B}) \right\rbrace - \lambda N \sumj \sum_{r=1}^R\, \vbar \gamma_{jr} \vbar,
$$
where $\mathbf{V}$ is the (orthogonal) $J\times R$ matrix of loadings $(R < J)$ with entries $\gamma_{jr}$ and $\mathbf{B}$ is a diagonal covariance matrix. Therefore, variables whose estimated loadings are $\gamma_{jr} = 0 \, \forall\, r$ will be non influential to the clustering and discarded. 

A related method is suggested by \cite{bouveyron:brunet:2014}, who propose to perform selection of the relevant variables by inducing sparsity in the loading matrix of the Fisher-EM algorithm (see \citep{bouveyron:2012,bouveyron:2014} for details). The method assumes that the observed data lie in a discriminative latent space  defined by latent features which are linear combinations of the original variables. As before, the density of each data point is expressed by:
$$
p(\x_i; \bm{\Theta}) = \sumg \tau_g \, \phi(\x_i ; \mathbf{V}\MU_g, \mathbf{V}\SIG_g\hspace*{-2pt}\mathbf{V}' \,+\, \mathbf{B}).
$$
The main idea is to obtain a sparse estimate of $\mathbf{V}$ such that variables whose loadings are shrunk to zero will not be relevant in defining the discriminative latent subspace and hence not useful for clustering. The authors propose three different approaches. The simplest, although easy to implement and competitive with the others, aims at obtaining a sparse approximation $\tilde{\mathbf{V}}$ of the estimate $\hat{\mathbf{V}}$ computed during an iteration of the Fisher-EM. To this purpose, the following minimization problem is posed:
$$
\tilde{\mathbf{V}} = \underset{\mathbf{V}}{\operatorname{argmin}} \left\lvert\left\lvert \X' \hat{\mathbf{V}} - \X' \mathbf{V}  \right\lvert\right\lvert_F \,+\, \lambda \sum_{r=1}^R \,\lvert\lvert \mathbf{v}_r \lvert\lvert_1
$$
where $\mathbf{v}_r$ is a column of $\mathbf{V}$ and $\lvert\lvert \cdot \lvert\lvert_F$ denotes the Frobenius norm. The other two are more involved and estimate a matrix $\mathbf{V}$ that is directly sparse and defines a discriminative subspace; we point to the referenced work for details.

For all the methods described above, maximization of the penalized likelihood for different specifications of $Q_{\lambda}(\cdot)$ is usually carried out resorting to a form of penalized EM algorithm \citep{green:1990}. Moreover, model selection consists in the selection of the number of mixture components and of the optimal shrinkage parameter $\lambda$, and to accomplish the task a modified BIC is employed; see the referenced works for details.

A method that differs in some way from the general form of \eqref{eq:2} is the \emph{sparse k-means} proposed by \cite{witten:tibshirani:2010}. Let $\mathbb{C} = \lbrace C_1,\, \dots,\, C_g,\, \dots,\, C_G \rbrace$ be a partition of the observations into disjoint subsets, and let $\mathbf{w} = (w_1,\, \dots,\, w_j,\, \dots,\, w_{J})$ be a vector of weights for each variable. Clustering and variable selection is performed by solving the optimization problem:
\begin{align*}
\underset{\mathbb{C},\, \mathbf{w},\, \MU}{\operatorname{argmax}} & \left\lbrace \sumj w_j \left[\sumi (x_{ij} - \bar{x}_j)^2 - \sumg \sum_{i\in C_g} (x_{ij} - \mu_{gj})^2 \right] \right\rbrace\\
\text{subject to}&\quad \lvert\lvert\,\mathbf{w}\,\lvert\lvert^2 \leq 1, \quad \lvert\lvert\,\mathbf{w}\,\lvert\lvert_1 \leq s,\\
\text{with}& \quad w_j \geq 0 \quad \forall j,
\end{align*}
where $\bar{x}_j$ is the sample mean of variable $j$ and $s$ is a tuning parameter. The $L_2$ penalty will force the weights in the interval $[0,1]$, while the $L_1$ norm is used to shrink the values to 0. The weight $w_j$ is interpreted as the contribution of $X_j$ to the resulting clustering: a large value of $w_j$ indicates a variable useful for clustering, while $w_j = 0$ means that variable $j$ does not contribute to the clustering. Note that the above objective function is related to the log-likelihood of a Gaussian mixture model with a common isotropic covariance matrix of the form $\SIG_g=\sigma^2\mathbb{I}$, an implicit assumption of the method. Indeed, in this exposition we only considered the case associated to Gaussian mixture modeling and we adapted equation (10) of \cite{witten:tibshirani:2010} to the purpose, using the connection of $k$-means to the fitting of mixture of Gaussians with common spherical covariance matrix (see the cited work for additional details). Furthermore, the method needs the number of clusters $G$ to be known in advance.  However, the approach is particularly suitable for high-dimensional data and the authors describe a more general framework for sparse clustering. Related work on sparse $k$-means is in \cite{sun:2012}, where consistency properties are investigated.

\subsection{Model selection approaches}
\label{vsgmm_modsel}
Within this class of approaches, the problem of selecting the relevant clustering variables is recast as a model selection problem. Different models are specified by the role attributed to the variables in connection to their relation with the clustering variable $\z$. Then these models are compared by means of a model selection criterion and relevant clustering variables are chosen accordingly to the best model.

\begin{figure}[!tb]
  \centering
    \includegraphics[scale=1]{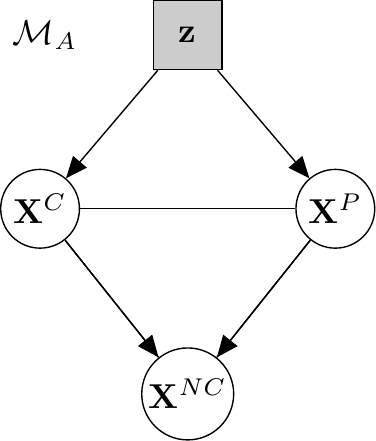}\quad\quad\includegraphics[scale=1]{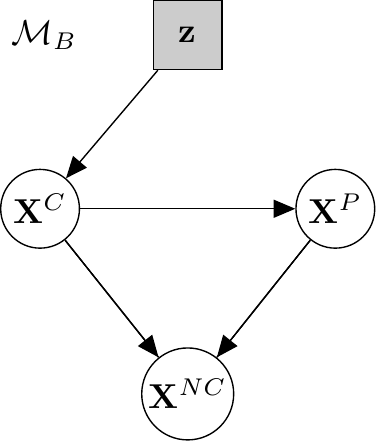}
  \caption{\label{graph1} The two competing models proposed in \cite{raftery:dean:2006}.}
\end{figure}

The framework was pioneered by the work of \cite{raftery:dean:2006}, where three main roles are specified for the variables. The authors propose a procedure where $\X$ is partitioned into the following subsets of variables:
\begin{itemize}[noitemsep]
 \item $\X^{C}$, the set of current clustering variables;
 \item $\X^{P}$, the variable(s) proposed to be added or removed from the set of clustering variables;
 \item $\X^{NC}$, the set of other variables not relevant for clustering.
\end{itemize}
Then the decision for inclusion or exclusion of $\X^P$ is taken by comparing the models (Fig~\ref{graph1}):
\begin{align*}
\mathcal{M}_A\colon \quad p(\X \vbar \z) &= p(\X^{C}, \X^{P} \vbar \z)\,p(\X^{NC}\vbar \X^C, \X^P),\\
\mathcal{M}_B\colon  \quad p(\X \vbar \z) &= p(\X^{C} \vbar \z)\,p(\X^P \vbar \X^C) \, p(\X^{NC}\vbar \X^C, \X^P).
\end{align*}
In model $\mathcal{M}_A$, $\X^P$ is useful for clustering and the joint distribution $p(\X^{C}, \X^{P} \vbar \z)$ corresponds to a Gaussian mixture distribution; on the other hand, $\mathcal{M}_B$ states that $\X^P$ does not depend on the clustering $\z$ and the conditional distribution $p(\X^P \vbar \X^C)$ corresponds to a linear regression. An important feature of the framework formulation is that in $\mathcal{M}_B$ the irrelevant variables are not required to be independent of the clustering variables. This allows to discard redundant variables related to the clustering ones but not to the clustering itself. Another important characteristic of the model construction lies in the avoidance of the unrealistic assumption of the independence between clustering and irrelevant variables. The competing models are compared using the BIC approximation to their marginal likelihoods:
\begin{align*}
\text{BIC}_A &= \text{BIC}_{\text{clust}}(\X^{C}, \X^{P}),\\
\text{BIC}_B &= \text{BIC}_{\text{no clust}}(\X^{C}) + \text{BIC}_{\text{reg}}(\X^{P} \vbar \X^C),
\end{align*}
where $\text{BIC}_{\text{clust}}(\X^{C}, \X^{P})$ is the BIC of a GMM in which $\X^P$ adds useful information to the clustering, $\text{BIC}_{\text{no clust}}(\X^{C})$ is the BIC of the GMM on the current set of clustering variables and $\text{BIC}_{\text{reg}}(\X^{P} \vbar \X^C)$ is the BIC of the regression of $X^P$ on $X^C$. Then if the difference $\text{BIC}_A - \text{BIC}_B$ is greater than zero, $\X^P$ is added to the set of clustering variables. To perform the selection, variables are added/removed and different models are compared using a stepwise algorithm.

\begin{figure}[!tb]
   \centering
    \includegraphics[scale=1]{graph_1.pdf}\quad\quad\includegraphics[scale=1]{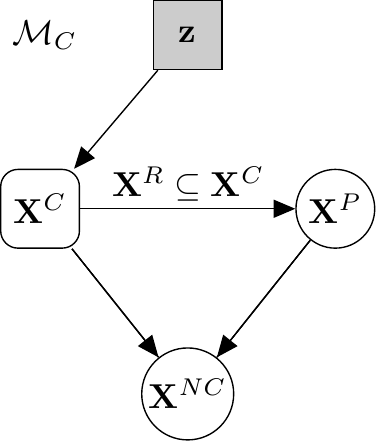}
   \caption{\label{graph2} The two models compared in \cite{maugis:celeux:2009:a}.}
\end{figure}  

The method has been further extended by \cite{maugis:celeux:2009:a} and \cite{maugis:celeux:2009:b}. \cite{maugis:celeux:2009:a} build on the modeling framework of \cite{raftery:dean:2006}, however, in the conditional distribution $p(\X^P \vbar \X^C)$, $\X^P$ can be related only to a subset $\X^R \subseteq \X^C$ of the clustering variables. Therefore, in the regression only a subset $\X^R$ of predictors are used to describe the dependency between $\X^P$ and $X^C$. In this way, the authors avoid the inclusion of unneeded parameters that would over-penalize the log-likelihood with the consequence of favoring models that declare some irrelevant variables as relevant when model comparison is performed using the BIC. The models compared in \cite{maugis:celeux:2009:a} are depicted in Fig.~\ref{graph2}, with $\mathcal{M}_A$ the same as previously stated and $\mathcal{M}_C$ defined as follows:
$$
\mathcal{M}_C\colon  \quad p(\X \vbar \z) = p(\X^{C} \vbar \z)\,p(\X^P \vbar \X^R \subseteq \X^C) \, p(\X^{NC}\vbar \X^C, \X^P),
$$
with $p(\X^P \vbar \X^R \subseteq \X^C)$ the regression term of $\X^P$ on the relevant predictors $\X^R$. For this model the BIC is given by:
$$
\text{BIC}_C = \text{BIC}_{\text{no clust}}(\X^{C}) + \text{BIC}_{\text{reg}}(\X^P \vbar \X^R \subseteq \X^C),
$$
with $\text{BIC}_{\text{reg}}(\X^P \vbar \X^R \subseteq \X^C)$ the BIC of the regression of $\X^P$ on $\X^C$ after selection of the optimal set of predictors $\X^R$. Again, to decide if $\X^P$ is relevant for clustering, the difference $\text{BIC}_A - \text{BIC}_C$ is computed and $\X^P$ is added to the set of clustering ones if this quantity is greater than zero. Model search is performed using a backward search and for selecting the relevant predictors in the regression the authors propose a standard stepwise procedure.

\begin{figure}[!tb]
  \centering
   \includegraphics[scale=1]{graph_1.pdf}\quad\quad\includegraphics[scale=1]{graph_3.pdf}\quad\quad\includegraphics[scale=1]{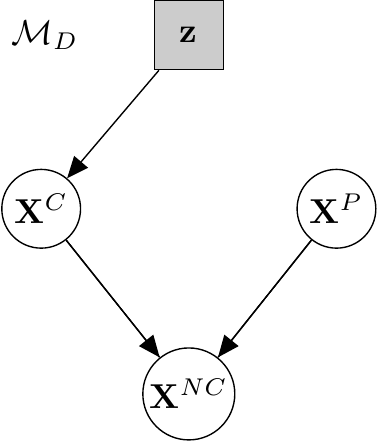}
  \caption{\label{graph3} The three different models presented in \cite{maugis:celeux:2009:b}.}
\end{figure}

The framework is subsequently expanded in \cite{maugis:celeux:2009:b}, where the authors consider an additional role for the variables proposed to be added or removed. They explicitly account for the case where $\X^P$ can be independent of the clustering variables $\X^C$. The assumption leads to further circumvent the over-penalization problem of the log-likelihood when parsimonious Gaussian mixture models are involved in the comparison and penalized likelihood criteria are used for model selection. The authors suggest a more flexible framework where three different models are compared, all represented in Fig.~\ref{graph3}. Models $\mathcal{M}_A$ and $\mathcal{M}_C$ are as before, while model $\mathcal{M}_D$ is specified as follows:
$$
\mathcal{M}_D\colon  \quad p(\X \vbar \z) = p(\X^{C} \vbar \z)\,p(\X^P) \, p(\X^{NC}\vbar \X^C, \X^P),
$$
with $p(\X^P)$ the density of a multivariate Gaussian distribution. Then, the corresponding BIC is given by:
$$
\text{BIC}_D = \text{BIC}_{\text{no clust}}(\X^{C}) + \text{BIC}_{\text{indep}}(\X^P),
$$
where $\text{BIC}_{\text{indep}}(\X^P)$ is the BIC of a multivariate Gaussian model. For an efficient model search, the authors suggest to rewrite $\text{BIC}_D$ as:
$$
\text{BIC}^*_D = \text{BIC}_{\text{no clust}}(\X^{C}) + \text{BIC}_{\text{reg}}(\X^P \vbar \tilde{\X}^R \subseteq \X^C),
$$
where $\tilde{\X}^R$ is allowed to be the empty set $\varnothing$ and thus $\text{BIC}_{\text{reg}}(\X^P \vbar \tilde{\X}^R \subseteq \X^C) = \text{BIC}_{\text{indep}}(\X^P)$ when $\tilde{\X}^R = \varnothing$. As in \cite{maugis:celeux:2009:a}, the difference $\text{BIC}_A - \text{BIC}^*_D$ is computed at each step of the algorithm and $\X^P$ is added to $\X^C$ if this quantity is positive. In the algorithm the variables are added and removed using a stepwise searching method and relevant predictors are selected by a backward stepwise algorithm. \cite{maugis:2012} extend the variable selection framework to handle data with missing values without resorting to any imputation procedure.

Note that the mentioned methods not only perform the selection of the relevant clustering variables, but at the same time they select the optimal number of clusters and the best parsimonious Gaussian clustering model from the family of models of \cite{celeux:govaert:1995}; we refer to the cited works for details. Lastly, it is worth mentioning two further extensions of the above modeling frameworks: \cite{scrucca:2016} proposes the use of genetic algorithms for searching over the whole model space to overcome the sub-optimality of a stepwise search, while \cite{galimberti:2017} suggest a framework where the relevant variables can provide information about more than one clustering structure of interest.

In the above approaches, at each step of the variable selection algorithm, the likelihood of a GMM has to be optimized several times. To deal with this computational issue, \cite{marbac:sedki:2017} propose a procedure that relies on the ICL criterion and does not require multiple calls of the EM algorithm. In the framework, a variable is declared as irrelevant to the clustering if its one-dimensional marginal distributions are equal between the mixture components. The authors introduce a binary indicator variable $\bm{\omega} = ( \omega_1,\, \dots,\, \omega_j,\, \dots,\, \omega_J )$ such that $\omega_j=1$ if variable $X_j$ is relevant, 0 otherwise. In this context we would denote $\X^C = \lbrace X_j \in \X \colon \omega_j = 1,\, \forall\, j \rbrace$ and $\X^{NC} = \lbrace X_j \in \X \colon \omega_j = 0,\, \forall\, j \rbrace$, and different models are specified by different configurations of the binary vector $\bm{\omega}$. These models are compared by means of a criterion based on the following integrated complete-data likelihood:
$$
p(\X, \z \vbar \bm{\omega}) = \int p(\X, \z \vbar \bm{\Omega}, \bm{\omega})\, p(\bm{\Omega} \vbar \bm{\omega} ) \,d \bm{\Omega},
$$
where $\bm{\Omega}$ denotes the collection composed of the parameters related to the mixture distribution for $\X^C$ and the parameters related to the distribution for $\X^{NC}$. After assuming local and global independence among the variables and placing conjugate priors on the parameters, the above integral reduces to:
$$
p(\X, \z \vbar \bm{\omega}) = p(\z)\prod_{j=1}^J p(X_j \vbar \omega_j, \z),
$$
where all the quantities have closed analytical expressions (see authors' paper for more details). For a non relevant variable, $p(X_j \vbar \omega_j, \z) = p(X_j \vbar \omega_j)$ since this quantity does not depend on the data classification, while for a relevant one this quantity will assume different values over the mixture components. Consequently, variable selection is performed by finding the vector $\bm{\omega}$ that maximizes the Maximum Integrated Complete-data Likelihood (MICL) criterion, given by:
$$
\text{MICL}(\bm{\omega}) = \log p(\X, \z_{\bm{\omega}}^* \vbar \bm{\omega}),
$$
with $\z_{\bm{\omega}}^* = \operatorname{argmax}_{\z} \log p(\X, \z \vbar \bm{\omega})$. Maximization of $\text{MICL}(\bm{\omega})$ is carried out iteratively using an algorithm that alternates between two optimizations steps of the ICL: optimization on the classification $\z$ given the data and $\bm{\omega}$, and maximization on $\bm{\omega}$ given the data and the classification $\z$. As in \cite{raftery:dean:2006}, \cite{maugis:celeux:2009:a} and \cite{maugis:celeux:2009:b}, the algorithm also returns the optimal number of clusters and we refer to the paper for details. The approach is fast and scales well for problems with a large number of variables. However, the optimization on $\z$ can be computationally demanding for large sample sizes; the authors suggest that for sample sizes smaller than $10^4$ this optimization is still doable. Furthermore, in a situation where variables are correlated and many redundant ones are present, the local and global independence assumptions could be too restrictive.

\subsection{Other approaches}
\label{vsgmm_other}
Other variable selection methods that do not univocally fit in the previous sections have been developed in the literature. Within a wrapper approach, \cite{dy:brodley:2004} propose to embed in the EM a forward selection algorithm for the maximization of two simple alternative criteria for variable selection. The first is based on a measure of separability and unimodality of the clusters, and consists in the quantity $\text{tr}(S_W^{-1} S_B)$, where $S_W$ is the within-cluster scatter matrix and $S_B$ is the between class scatter matrix. The second is based on the quantification of how well the model fits the data and the natural choice is the likelihood of a GMM itself.

A related method is the one of \cite{andrews:2014}. The authors suggest an hybrid filter-wrapper approach based on a variable within-group variance, given by: 
$$
W_j = \dfrac{\sumg \sumi z_{ig} (x_{ij} - \mu_j)^2}{N}.
$$
First, an initial estimate of $W_j$ is found by running a preliminary clustering step and the variables are listed in ascending order according to their values. The top variable in the list is automatically selected (as the one with the minimum $W_j$) and placed in the set of selected variables $\mathcal{S}$. Subsequently, a variable $j$ is selected if
$$
\vbar \rho_{rj} \vbar\, < \,(1 - W_j)^c \quad \forall \,\, X_r \in \mathcal{S},
$$
with $\rho_{rj}$ the correlation between $X_j$ and $X_r$, and $c$ a coefficient used to weight the within group variation. The authors suggest to use the integer values from 1 to 5 and the procedure will tend to include more variables in the model as $c$ increases. Lastly, clustering is performed by fitting a GMM on the final set $S$. Note that before running the procedure the data need to be standardized to have mean 0 and variance 1. 

Another method using the concept of separability and unimodality of the clusters as in \cite{dy:brodley:2004} is the one found in \cite{lee:li:2012}. Here GMMs are used to estimate the density of the data and the variable selection is based on a cluster separability measure defined using ridgelines, one-dimensional parametric curves linking the modes of two clusters as function of their densities. Let $y_{gk}(a_t)$, be the set of points individuating the ridgeline between cluster $g$ and $k$ ($a_0 = 0 < a_1 < \dots < a_T = 1, t = 0,\,\dots,\, T)$;  the pairwise separability between these clusters is measured by:
$$	
1- \dfrac{\text{min}_{t=1}^T\left\lbrace \sum_{h=1}^G \phi(y_{gk}(a_t); \MU_h, \SIG_h) \right\rbrace}{\text{min}\left\lbrace \sum_{h=1}^G \phi(y_{gk}(0); \MU_h, \SIG_h),\, \sum_{h=1}^G \phi(y_{gk}(1); \MU_h, \SIG_h) \right\rbrace}.
$$
The aim is to find the set of variables that achieves the maximum aggregated separability across all the pairs of clusters and the authors suggest two alternative forward selection algorithms for the task. The method tends to select variables indicating well-separated clusters.

A simple filter approach used to pre-select the variables and reduce the dimensionality of the data is the one suggested by \cite{mclachlan:bean:2002}, within the context of clustering microarray expression data. In this approach, first a univariate mixture model is fitted to each variable in turn. Then the likelihood ratio statistic for the test of a single normal distribution versus a mixture of Gaussians is computed and those variables for which this statistic is significant are retained. The selected variables are subsequently clustered using a k-means procedure in order to find those centroids that represent subsets of variables. Model-based clustering is consequently performed on these representative centroids, which represent the data in a lower dimensional subspace. 

The stepwise algorithm presented in \cite{maugis:celeux:2009:b} could be very slow in high-dimensional settings. To overcome the computational burden, \cite{celeux:2017} suggest an hybrid approach where a LASSO-like procedure and the model selection algorithm are employed together. The approach consists of two subsequent steps. First, the method of \cite{zhou:2009} is applied to the data (see Section~\ref{vsgmm_pen}) and a ranking of the variables is obtained. For fixed $G$ and a set of different combinations $L_{\lambda_1}\times L_{\lambda_2}$ of values $(\lambda_1,\lambda_2)$, the \emph{clustering score} of each variable $X_j$ is defined as:
$$
\mathcal{O}_G(j) = \sum_{L_{\lambda_1}\times L_{\lambda_1}} \mathcal{B}_{G,\lambda_1,\lambda_2}(j),
$$
where
$$
\mathcal{B}_{G,\lambda_1,\lambda_2}(j) = \begin{cases}
					     0	\quad \text{if}\quad \hat{\mu}_{1j}(\lambda_1,\lambda_2) = \, \dots\, =\hat{\mu}_{gj}(\lambda_1,\lambda_2)=\, \dots\, =\hat{\mu}_{Gj}(\lambda_1,\lambda_2) = 0,\\
					     1 \quad \text{if}\quad \hat{\mu}_{gj}(\lambda_1,\lambda_2) \neq 0, ~ g \in \lbrace 1,\, \dots,\, G \rbrace. 
					 \end{cases}
$$
The larger the value of $\mathcal{O}_G(j)$, the more a variable is likely to be relevant for the clustering. Then the variables are ranked in decreasing order according to this quantity; denote this rank as $\mathcal{I}_G$. Ideally, in the rank relevant variables are in the top positions, then redundant variables and non-informative ones at the end. In the second step, \cite{maugis:celeux:2009:b} method is applied using this rank. The variable set is scanned according to the $\mathcal{I}_G$ order and variables are declared as relevant until for $c$ consecutive variables there is no evidence of being clustering variables. Then irrelevant variables are determined scanning the set in the reverse order of $\mathcal{I}_G$. The process stops as soon as $c$ consecutive variables are not declared as non-informative. The redundant variables are thus determined by the remaining variables. The tuning parameter $c$ is a buffer to reduce the effect of mistakenly ranked variables in the LASSO-like step.

Another hybrid contribution is the work of \cite{malsiner:etal:2016}. The authors define a full Bayesian framework for estimation of sparse Gaussian mixtures where the assumption of a local shrinkage prior on the component means allows the identification of cluster-relevant variables. They suggest a Normal-Gamma prior of the form:
$$
\mu_{gj} \vbar \lambda_j, b_{0j} \sim \mathcal{N}(b_{0j}, \lambda_j s_0),
$$
where
$$
\lambda_j \sim \mathcal{G}(\nu_1, \nu_2), \quad b_{0j} \sim \mathcal{N}(m_0, M_0),
$$
with $\nu_1, \nu_2, m_0, M_0$ and $s_0$ some hyperparameters. For values of $\nu_1 < 1$, the prior shrinks all the component means of variable $j$ towards the common value $b_{0j}$, with the effect that variables uninformative for the clustering are effectively fit by a single mixture component. Note that this approach could be interpreted as an hybrid form of Bayesian and penalized variable selection procedures.

Although not directly related to the Gaussian mixture modeling framework, it is worth to mention two other approaches that can be thought of a cross-over between Bayesian and penalization methods. The first is the work from \cite{hoff:2006}, where it is proposed a Bayesian framework for finding subsets of variables that distinguish the clusters from each other. Thus, every cluster is characterized by its own set of discriminating variables that differentiates it from the rest of the clusters. In the method, each data point is modeled as:
$$
x_{ij} = \bar{\mu}_j + \varphi_{gj}\, \mu_{gj}\, \sigma_{gj} + \sigma_{gj}^{\varphi_{gj}}\, \epsilon_{ij},
$$
where $\bar{\mu}_j$ is the overall data mean of variable $j$, $\varphi_{gj}$ is a binary variable indicating if $X_j$ is ``active'' on cluster $g$ and $\epsilon_{ij}$ is an error component such that $\epsilon_{ij} \sim \mathcal{N}(0, \bar{\sigma}^2_j)$, with $\bar{\sigma}^2_j$ the overall data variance of $X_j$. The model results in a multivariate Dirichlet process mixture model and the use of conjugate priors  allows the implementation of a Gibbs sampling scheme for inference. Rather than performing a genuine variable selection, the method identifies the cluster-specific relevant variables by detecting the shifts in means and variances from the corresponding overall data quantities. The approach belongs to the class of \emph{subspace clustering} methods and is related to the work of \cite{friedman:2004}.

The second is the one proposed by \cite{partovi:2015} and is closely related. In this case, the authors consider the following linear model for the data points:
$$
x_{gij} = \bar{\mu}_j + \gamma_j \, \varphi_{gj}\, \mu_{gj} + \epsilon_{gij},
$$
where $x_{gij}$ is the data of clustering observation $i$ measured on variable $j$ in cluster $g$; $\gamma_j$ and $\varphi_{gj}$ are binary variables such that $\gamma_j\sim\text{Bernoulli}(\nu)$ and $\varphi_{gj}\sim\text{Bernoulli}(\eta)$ respectively. Probability $\nu$ may be interpreted as the prior proportion of relevant variables, while $\eta$ as the prior proportion of non-overlapping component means for a clustering variable. Two families of prior distributions are suggested for $\mu_{gj}$: Gaussian and asymmetric Laplace. In both cases, the marginal distribution of the observations is given in closed form and results in a \emph{spike and slab} density in which the data is modeled by a mixture of two densities. When $(\gamma_j = 1, \varphi_{gj} = 1)$, the variable is active and the data is modeled by the slab density which drives the clustering procedure when a variable is relevant for clustering. When $\gamma_j = 0$ or $\varphi_{gj} = 0$, the density of the data corresponds to the spike density, which reduces the effect of non-informative variables. Variables not important for any of the clusters can be discarded. The setting allows the use of Bayes factors for computing the importance of each variable and perform the selection, and clustering of the data is achieved via an agglomerative hierarchical procedure.

\subsection{R packages and data example}
The available \textsf{R} packages for variable selection for Gaussian mixture models are: \texttt{sparcl} \citep{sparcl}, \texttt{clustvarsel} \citep{scrucca:raftery:2015}, \texttt{VarSelLCM} \citep{varsellcm}, \texttt{vscc} \citep{vscc}, \texttt{SelvarMix} \citep{selvarmix}, and \texttt{bclust} \citep{partovi:2012}. Table~\ref{pack_GMM} lists them, with information about the method and the type of approach. It is also worth to mention the package \texttt{ClustOfVar} \citep{clustofvar}. Rather than performing variable selection and obtain a classification of the data points, this package aims to find clusters of variables linked to a central synthetic variable obtained from a principal components decomposition of the data; for this reason, it will not be considered in the subsequent analysis. Lastly, we also point to the C++ softwares \texttt{SelvarClust}\footnote{\url{http://www.math.univ-toulouse.fr/~maugis/SelvarClustHomepage.html}} and \texttt{SelvarClustIndep}\footnote{\url{http://www.math.univ-toulouse.fr/~maugis/SelvarClustIndepHomepage.html}}, which implement the modeling framework of \cite{raftery:dean:2006} and \cite{maugis:celeux:2009:a,maugis:celeux:2009:b}. We will not consider these software in the following analysis, since the aim of this paper is in reviewing \textsf{R} packages.

\begin{table}[!t]
\centering
\caption{\label{pack_GMM} \textsf{\em R} packages for GMMs variable selection.}
{
 \begin{tabular}{lll}
 \toprule
  \bf Package	&	\bf Type	&	\bf Method\\
  \midrule
  \texttt{sparcl}	&	penalized	&	\cite{witten:tibshirani:2010}\\[1em]
  \multirow{3}{*}{\texttt{clustvarsel}}		&	\multirow{3}{*}{model selection}	&	\cite{maugis:celeux:2009:a}\\
  				&				&	\cite{maugis:celeux:2009:b}\\
  				&				&	\cite{scrucca:raftery:2015}\\[1em]
  \texttt{VarSelLCM}	&	model selection	&	\cite{marbac:sedki:2017}\\[1em]
  \texttt{vscc}		&	filter/wrapper		&	\cite{andrews:2014}\\[1em]
  \multirow{2}{*}{\texttt{SelvarMix}}	&	\multirow{2}{*}{model selection/penalized}	&	\cite{celeux:2017}\\
					&							&	\cite{sedki:2014}\\[1em]
  \texttt{bclust}	&	other/Bayesian	&	\cite{partovi:2015,partovi:2012}\\
  \bottomrule
 \end{tabular}%
 }
\end{table}

The adjusted Rand index \citep[ARI,][]{hubert:arabie:1985} will be employed to assess and compare the clustering performance of the packages. Let us consider two different partitions of the $N$ data points, one into $G$ clusters and the other into $K$ clusters; this index is defined as:
$$
\text{ARI} = \dfrac{ \sum^G_{g} \sum^K_{k} \binom{N_{gk}}{2} - \left[\sum^G_g \binom{N_{g.}}{2} \sum^K_k \binom{N_{.k}}{2}\right] / \binom{N}{2}}{ \frac{1}{2} \left[\sum^G_g \binom{N_{g.}}{2} + \sum^K_k \binom{N_{.k}}{2}\right] - \left[\sum^G_g \binom{N_{g.}}{2} \sum^K_k \binom{N_{.k}}{2}\right] / \binom{N}{2}},
$$
where $N_{gk}$ is the number of observations falling in cluster $g$ and cluster $k$, while $N_{g.}= \sum_k N_{gk}$ and $N_{.k}= \sum_g N_{gk}$. The ARI corrects the comparison for the fact that two partitions could match merely by chance \citep{hubert:arabie:1985}. The index has a maximum value of 1 and attains this value when there is perfect matching between the two partitions; on the other hand, it has expected value of 0 when the two partitions are completely independent. Therefore, the index is used to measure the agreement between two different classifications of the data.

\begin{figure}[!t]
  \centering
    \includegraphics[scale=0.5]{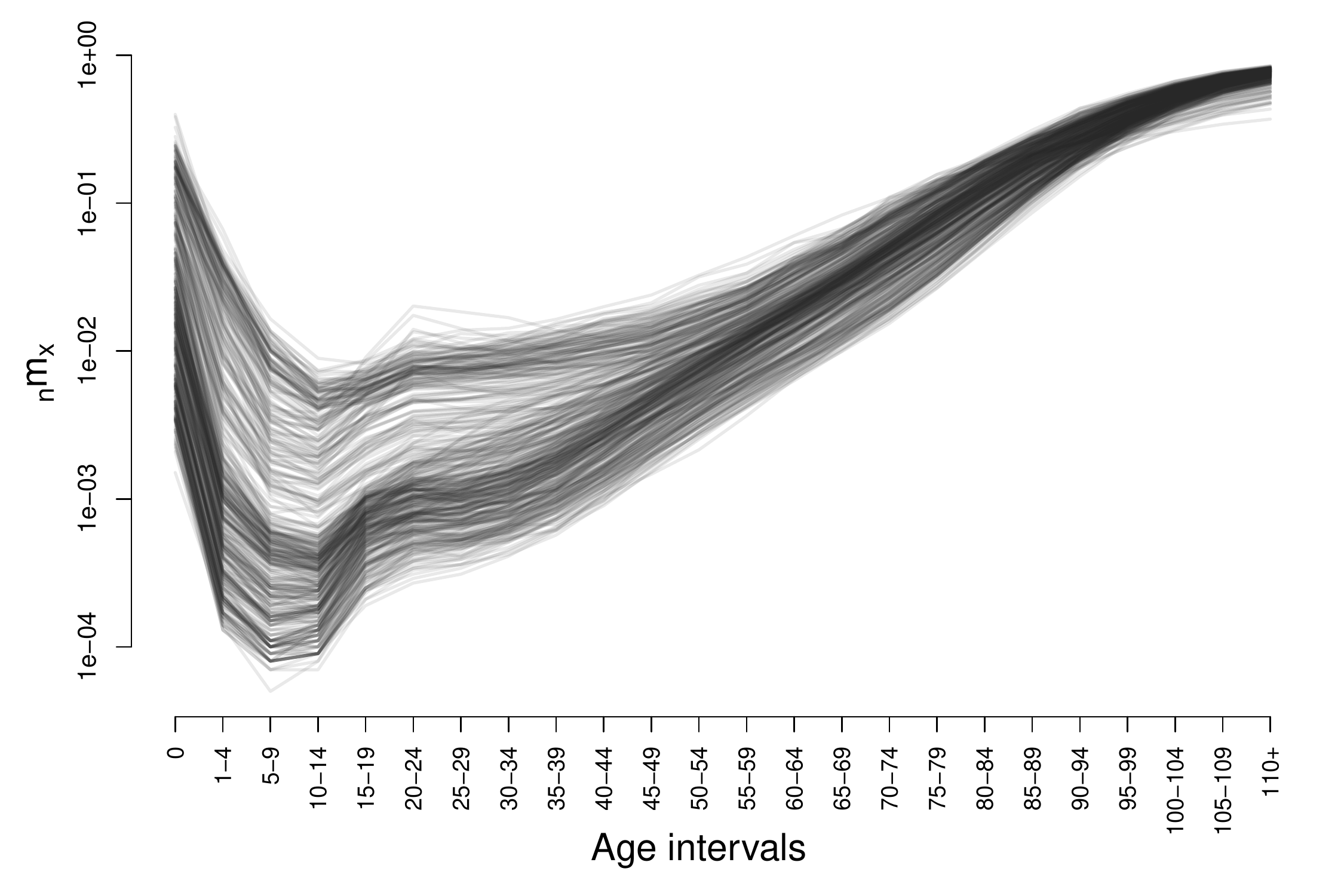}
  \caption{\label{hmd} Central mortality rates (values on logarithmic scale).}
\end{figure}

We apply the different variable selection methods to data from the Human Mortality Database \citep{hmd}. The database contains life tables from several countries, spanning a period of time from the middle of the 18\textsuperscript{th} century to 2015. We consider central mortality rates for both genders over 10-year time intervals and 5-year age intervals. The central mortality rate $_nm_x$ is interpreted as the average 10-year interval rate at which people die during the period between age $x$ and $x+n$, normalized by the number of those living. The data contain 389 age patterns of mortality, each over a 10-year period time for 24 age intervals from 0 to 110+ (Figure~\ref{hmd}). The aim is to cluster the mortality schedules and individuate those age groups relevant for the clustering structure. There are 27 non overlapping 10-year periods, some present multiple times (especially recent ones), and the patterns of mortality tend to cluster according to the historical period \citep{clark:2011}. To assess the quality of the clustering, we will compare the estimated partition with the information regarding the 10-year period each schedule belongs to using the ARI. 

\begin{figure}[!b]
\centering
  \includegraphics[scale=0.55]{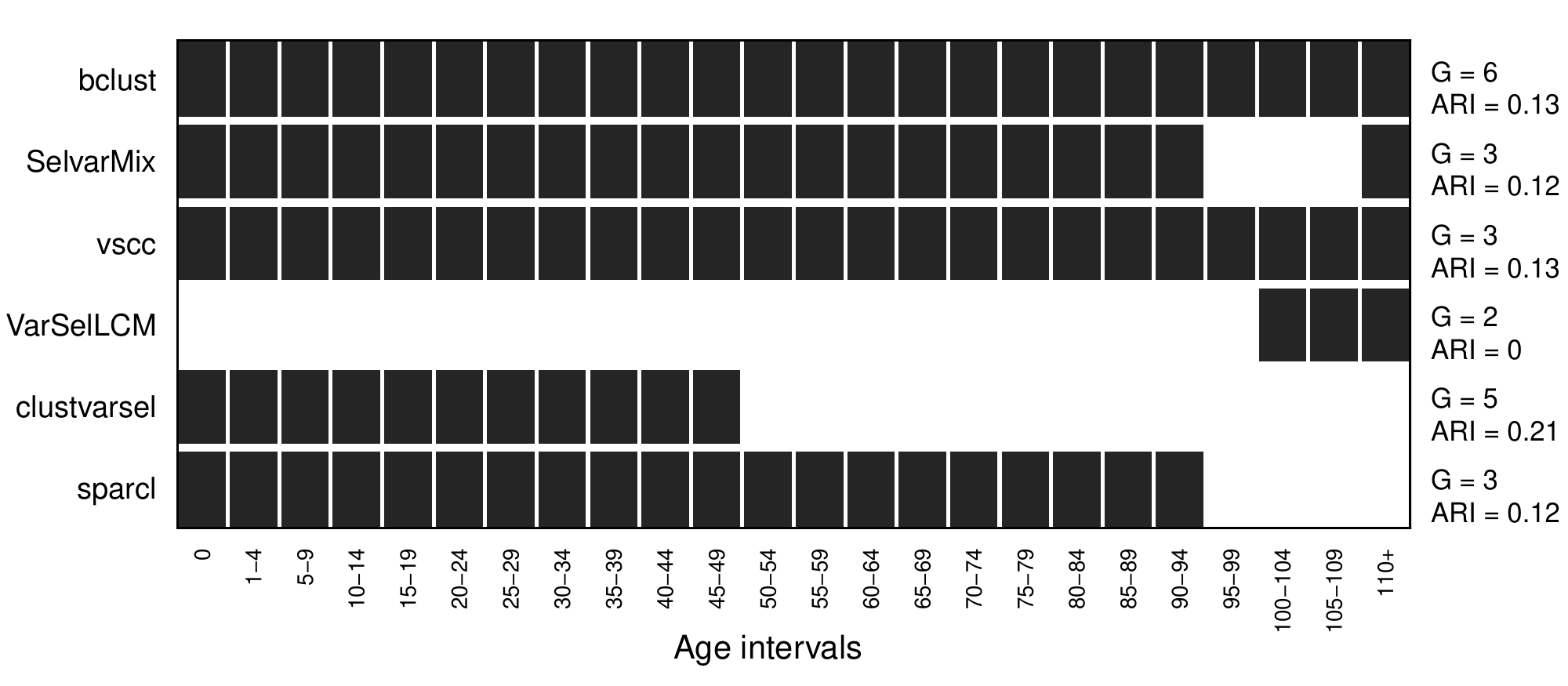}
  \caption{\label{fig_gmm_varsel_1} Variables selected via the \textsf{\em R} packages for GMMs variable selection listed in Table~\ref{pack_GMM}. A dark square indicates the variable has been selected. For each method, also the selected number of clusters and the ARI between the estimated classification and the 10-year historical period are reported.}
\end{figure}

\begin{figure}[!tb]
  \centering
    \includegraphics[scale=0.5]{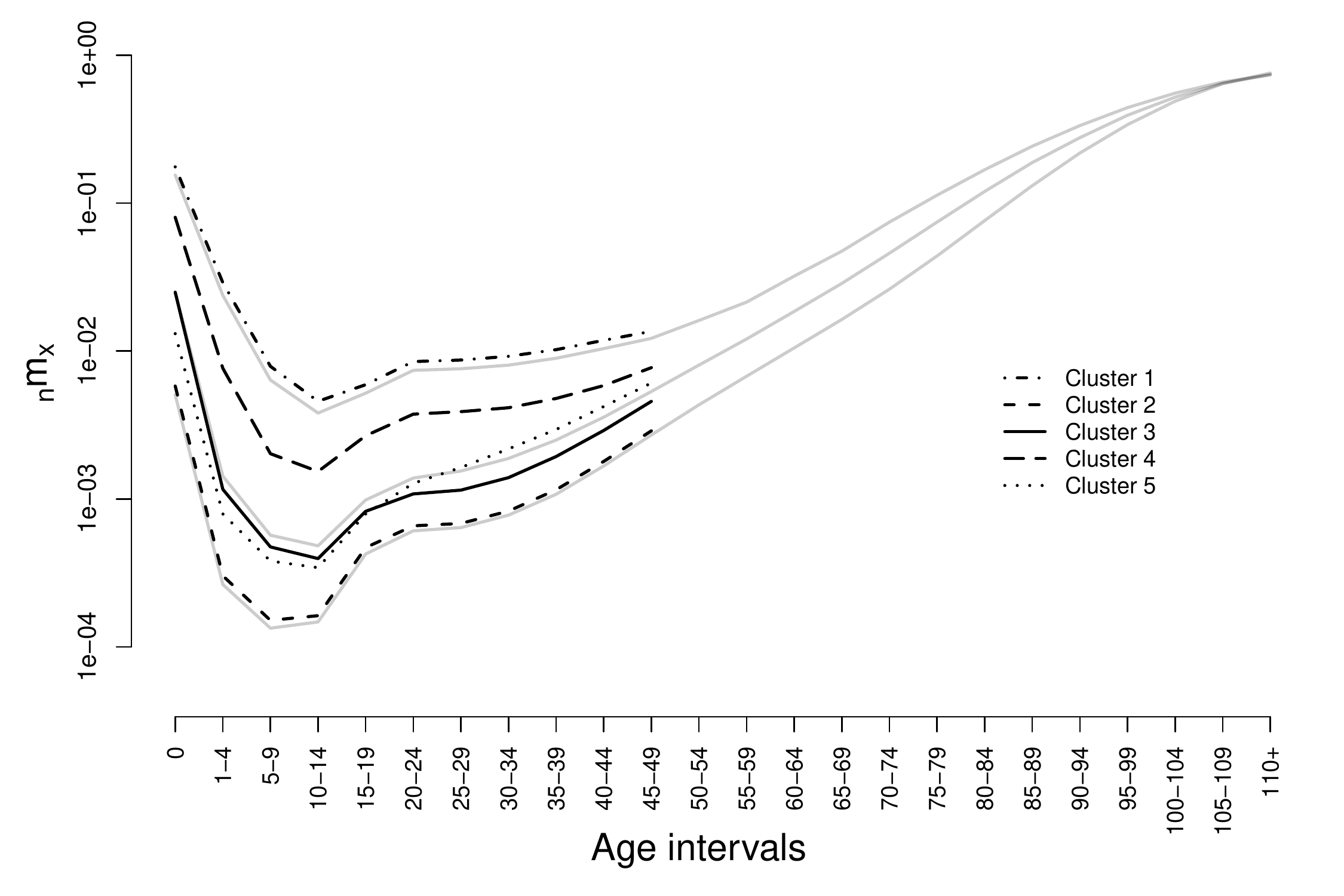}
  \caption{\label{fig_gmm_varsel_2} Average cluster mortality rate estimated by \texttt{\em clustvarsel} on the selected age groups. The full lines denote the \texttt{\em mclust} estimated cluster means on the whole data. Values are on logarithmic scale.}
\end{figure}

First we fit a Gaussian mixture model using the \texttt{mclust} package. The model selected is a 3-component mixture, with an ARI of 0.13 between the estimated classification and the historical period. Then we apply the variable selection methods listed in Table~\ref{pack_GMM}. Figure~\ref{fig_gmm_varsel_1} displays the selected age groups for each method, alongside the chosen number of clusters and the ARI between the estimated classification and the historical period. Note that for \texttt{sparcl} the number of components needs to be set in advance and we fixed it to the number selected by \texttt{mclust}. Package \texttt{bclust} did not discard any variables, most likely due to the fact that adjacent age groups are highly correlated and the independence assumption of this method is unrealistic in this case. Also package \texttt{vscc} did not remove any of the variables. Package \texttt{VarSelLCM} retained the smallest subset of variables. However, the package obtained an ARI of zero and surprisingly selected only the oldest age intervals, those expected to have the least information, thus it seems to have detected a spurious solution. We ran the package multiple times and the other best solutions found were a model with 3 clusters and the same subset of variables, and a model with only one cluster and no variables discarded. The package seems to be sensitive to the initialization and, also in this case, this is an indication that the strong independence assumptions at the basis of the method implemented could be too restrictive. The other packages, \texttt{sparcl}, \texttt{SelvarMix}, and \texttt{clustvarsel} discarded age groups representing the old population. Among these, \texttt{clustvarsel} performed the most parsimonious selection, retaining only 11 age groups in the range from 0 to 49 years old, and with the highest ARI. The clusters are discriminated along the young and middle ages, as it would be expected since for elder ages the differences among death rates level off. Table~\ref{cross-gmm} reports the cross-tabulation between the historical periods and the \texttt{clustvarsel} partition on the selected age intervals. The clusters have a meaningful interpretation in terms of the time dimension. Figure~\ref{fig_gmm_varsel_2} shows the estimated cluster means on the age groups selected by \texttt{clustvarsel}. The \texttt{mclust} estimated cluster means are added for comparison. In particular, the mortality patterns of clusters 1, 3 and 4 estimated on the selected variables are close to the average patterns estimated on all the data. The extra cluster 2 seems to capture the mortality pattern related to the period corresponding to the two World wars. Moreover, cluster 5 has a mean rate similar to cluster 3, but higher death rate for age intervals above 20 years old and captures the extra-heterogeneity present in the more recent years. 

\begin{table}[!t]
\centering
\caption{\label{cross-gmm} Cross-tabulation between the historical period classification and the \texttt{\em clustvarsel} partition on the selected age groups.}
\begin{tabular}{lrrrrrr}
  \toprule
 & & \multicolumn{5}{c}{Group}\\
 & &  1 & 2 & 3 & 4 & 5 \\ 
  \midrule
  \parbox[t]{2mm}{\multirow{27}{*}{\rotatebox[origin=c]{90}{10-year time period}}}&1750-1759 &   1 &    &   &    &    \\ 
  &1760-1769 &   1 &     &     &     &     \\ 
  &1770-1779 &   1 &     &     &     &     \\ 
  &1780-1789 &   1 &     &     &     &     \\ 
  &1790-1799 &   1 &     &     &     &     \\ 
  &1800-1809 &   1 &     &     &     &     \\ 
  &1810-1819 &   2 &     &     &     &     \\ 
  &1820-1829 &   2 &     &     &     &     \\ 
  &1830-1839 &   4 &     &     &     &     \\ 
  &1840-1849 &   6 &     &     &     &     \\ 
  &1850-1859 &   7 &     &     &     &     \\ 
  &1860-1869 &   7 &     &     &     &     \\ 
  &1870-1879 &  10 &     &     &     &     \\ 
  &1880-1889 &  10 &     &     &     &     \\ 
  &1890-1899 &   8 &   2 &     &     &     \\ 
  &1900-1909 &   7 &   4 &     &     &     \\ 
  &1910-1919 &   5 &   6 &     &     &     \\ 
  &1920-1929 &   3 &  11 &     &     &     \\ 
  &1930-1939 &     &  15 &     &     &     \\ 
  &1940-1949 &   4 &  15 &   1 &     &     \\ 
  &1950-1959 &     &   5 &  21 &     &   1 \\ 
  &1960-1969 &     &   3 &  25 &     &   6 \\ 
  &1970-1979 &     &     &  26 &   2 &   7 \\ 
  &1980-1989 &     &   1 &   8 &  19 &  10 \\ 
  &1990-1999 &     &     &   2 &  26 &  12 \\ 
  &2000-2009 &     &     &     &  34 &   7 \\ 
  &2010-2015 &     &     &     &  30 &   9 \\ 
  \bottomrule
\end{tabular}
\end{table}

Table~\ref{time-gmm} reports the computing time of the packages considered in Table~\ref{pack_GMM} on a Dell machine with Intel Core i7-3770 CPU @3.40GHz$\times$8. The packages are listed from the fastest to the slowest. \texttt{vscc} is the fastest among the packages considered, but it did not discard any of the age intervals. Package \texttt{sparcl} is the second fastest. In the case of highly correlated variables, \texttt{clustvarsel} performs the most parsimonious selection and obtains the best classification, with an acceptable computing time. Note, however, that in using \texttt{clustvarsel} we considered only 4 out of the 14 covariance models available, as some of them may require a long computational time to be fitted and to consider a set of models consistent with \texttt{SelvarMix}. Therefore, aside from \texttt{vscc}, \texttt{sparcl} is the fastest and as such it is particularly suitable to perform variable selection in high-dimensional settings where it is reasonable to assume local independence among the variables. Nevertheless, the package requires the number of clusters to be specified in advance, while \texttt{clustvarsel} and \texttt{SelvarMix} automatically select such number. \texttt{bclust} and \texttt{VarSelLCM} are the slowest in this example, and care may need to be taken when using these packages on highly correlated data, since they make use of independence assumptions that could be too restrictive. Lastly, we point that \texttt{clustvarsel}, \texttt{SelvarMix} and \texttt{VarSelLCM} allow for parallelization of the computations.

\begin{table}[!bt]
\centering
\caption{\label{time-gmm} Computing time (in seconds) and relative computing time of the R packages for variable selection for GMMs.}
\begin{tabular}{lrr}
\toprule
\bf Package	&	\bf Time (sec.)	&	\bf Relative \\
\midrule
\texttt{vscc} & 10 & 1.00 \\
\texttt{sparcl} & 42 & 4.09 \\ 
\texttt{clustvarsel} & 51 & 5.01 \\ 
\texttt{SelvarMix} & 61 & 5.90 \\ 
\texttt{VarSelLCM} & 438 & 42.58 \\ 
\texttt{bclust} & 916 & 89.14 \\ 
\bottomrule
\end{tabular}
\end{table}

\section{Variable selection methods for latent class analysis}
\label{vslca}
After the description of the main variable selection methods for GMMs, in this section we present the different approaches for variable selection in latent class analysis.

\subsection{Bayesian approaches}
Only recently, attention has been posed on the use of the Bayesian framework for variable selection in latent class analysis. The overall setting is similar to the one provided in Section~\ref{vsgmm_bayes} and we refer there for its description. Moreover, generally the methods borrow ideas from Bayesian variable selection for Gaussian mixture models.

Indeed, \cite{silvestre:cardoso:2015} propose an adaptation of \cite{law:figueiredo:2004} method to categorical data clustering and variable selection. The mixture of Gaussian densities is replaced by the mixture in \eqref{lca} and the authors use the same concept of feature saliency. Analogously, a Dirichlet distribution is assumed on the saliences and an EM algorithm for MAP estimation is employed.

\cite{white:wyse:2014} present an approach inspired by the work of \cite{tadesse:sha:2005}. The authors propose the of use a collapsed Gibbs sampler algorithm in which all the mixture parameters are integrated out. A prior of the form $p(\vphi \vbar \eta) = \prod_{j} \eta^{\varphi_j} (1-\eta)^{1-\varphi_j}$ is assumed for the indicator variable $\vphi$. Then, assuming conjugate priors on all the parameters, it is possible to marginalize them out analytically, obtaining the posterior:
$$
p(G, \z, \vphi \vbar \X) \propto p(G) \int p(\z, \vphi, \bm{\Omega} \vbar \X, G)\,d\bm{\Omega},
$$
with $\bm{\Omega}$ denoting the collection of parameters of the latent class analysis model and the parameters of the distribution for the irrelevant variables. The above quantity has a closed form expression and allows to implement a simple MCMC scheme. Subsequently, inference is performed using a post-hoc procedure. Variable selection and selection of $G$ is carried out jointly, by calculating the proportion of time the sampler spent in a certain number of groups and including a certain variable. The authors employ the method on different datasets and it is shown to outperform a more involved reversible jump MCMC algorithm for variable selection.

\subsection{Penalization approaches}
Despite of the large amount of work developed for penalized model-based clustering with GMMs (confront Section~\ref{vsgmm_pen}), little work has been done in the context of LCA. Additionally, more attention has been given to the aim of model identifiability and regularization, rather than to the purpose of variable selection per se. 

Within this framework enters the work of \cite{houseman:2006}. The authors propose a penalization approach for latent class regression analysis of high-dimensional genomic data where a set of categorical response variables is related to a set of numerical covariates. Here the class-conditional probabilities $\bm{\pi}_g$ in~\eqref{lca} are expressed as function of a vector of covariates and a set of regression coefficients. Then a ridge or LASSO penalty is used in order to obtain sparse estimates for the vectors of regression coefficients. Consequently, by discarding those predictors with coefficients shrunken to zero, the clusters would be characterized by different sets of selected covariates used to model the categorical responses. However, variable selection is not performed on the variables directly involved in the clustering.

\cite{wu:2013} presents a related approach, but this time focusing on the selection of variables involved in the clustering. In the method, the class conditional probabilities $\pi_{gjc}$ of observing category $c$ for variable $j$ within class $g$ are re-parameterized in terms of the logit transform as:
$$
\pi_{gjc} = \dfrac{\text{exp}(\alpha_{jc} + \beta_{gjc})}{\sum_{l=1}^{C_j}\text{exp}(\alpha_{jl} + \beta_{gjl})}, \qquad \text{with} \qquad \alpha_{j1} = \beta_{gj1} = 1.
$$
Then, similarly to \eqref{eq:2} in Section~\ref{vsgmm_pen}, the following penalized log-likelihood is considered:
$$
\ell_Q = \sumi \log \left\lbrace \sumg \tau_g \, \mathcal{C}(\x_i ; \bm{\alpha}, \bm{\beta}_g ) \right\rbrace - \lambda \sumg \sumj \sum_{c=1}^{C_j}\, \vbar \beta_{gjc} \vbar,
$$
where $\bm{\alpha}$ and $\bm{\beta}_g$ are the collections of parameters $\alpha_{jc}$ and $\beta_{gjc}$ respectively, and we made explicit the dependence on them of the Multinomial probability mass function. For a variable $X_j$, the quantities $\beta_{gjc}$ measure the difference from the overall probabilities $\text{exp}(\alpha_{jc})/\left[\sum_{l=1}^{C_j}\text{exp}(\alpha_{jl})\right]$ and its contribution to the clustering. Consequently, if $\sumg \sum_{c=1}^{Cj} \,\vbar \beta_{gjc}\vbar = 0$, variable $X_j$ is considered not influent to the classification and discarded. As for GMMs, the estimation in this setting is carried out by means of a penalized EM algorithm and the optimal $\lambda$ and the number of clusters $G$ are selected by BIC.

\subsection{Model selection approaches}
\begin{figure}[!tb]
   \centering
    \includegraphics[scale=1]{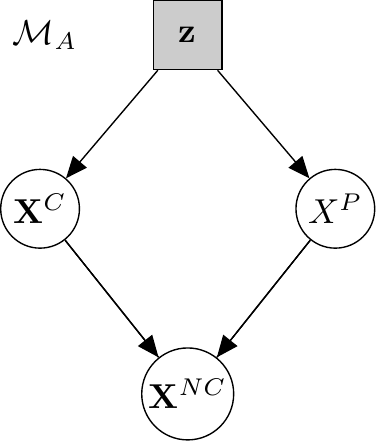}\quad\quad\includegraphics[scale=1]{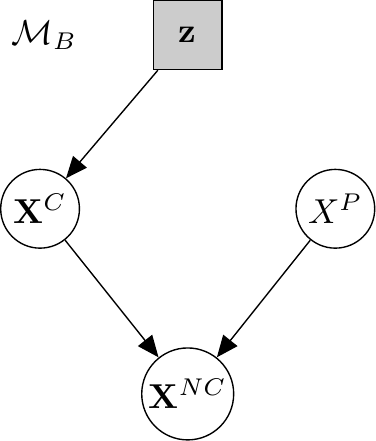}
   \caption{\label{graph4} The two models presented in \cite{dean:raftery:2010}.}
\end{figure}

Model selection approaches for variable selection in LCA draw from the work developed for GMMs. In \cite{dean:raftery:2010} the authors suggest a framework similar to the one presented in \cite{raftery:dean:2006}. As in section~\ref{vsgmm_modsel}, the following partition of $\X$ is considered:
\begin{itemize}[noitemsep]
 \item $\X^{C}$, the set of current clustering variables;
 \item $X^{P}$, the variable proposed to be added or removed from the set of clustering variables;
 \item $\X^{NC}$, the set of variables not relevant for clustering.
\end{itemize}
Then the authors propose to compare the two models represented in Fig.~\ref{graph4}. Model $\mathcal{M}_A$ states that $X^P$ is a clustering variable and there is no edge between $\X^{C}$ and $X^{P}$ due to the local independence assumption of LCA that implies $p(\X^C,X^P \vbar \z) = p(\X^C \vbar \z)p(X^P \vbar \z)$. In model $\mathcal{M}_B$, $X^P$ is not useful for clustering (the missing edge between $\z$ and $X^P$) and is assumed to be independent from the set $\X^C$. Under these models, the joint distribution of $\X$ is expressed as:
\begin{align*}
\mathcal{M}_A\colon \quad p(\X \vbar \z) &= p(\X^{C}, X^{P} \vbar \z)\,p(\X^{NC}\vbar \X^C, X^P),\\
\mathcal{M}_B\colon  \quad p(\X \vbar \z) &= p(\X^{C} \vbar \z)\,p(X^P) \, p(\X^{NC}\vbar \X^C, X^P),
\end{align*}
with $p(\X^{C}, X^{P} \vbar \z)$ the LCA model on on the clustering variables and $X^P$, $p(X^P)$ the Multinomial distribution and $ p(\X^{NC}\vbar \X^C, X^P)$ the distribution for the non-informative variables. The two models are compared via the BIC approximations:
\begin{align*}
\text{BIC}_A &= \text{BIC}_{\text{clust}}(\X^{C}, X^{P}),\\
\text{BIC}_B &= \text{BIC}_{\text{no clust}}(\X^{C}) + \text{BIC}(X^{P}),
\end{align*}
where $\text{BIC}_{\text{clust}}(\X^{C}, X^{P})$ is the BIC of a LCA model in which $X^P$ is a relevant clustering variable, $\text{BIC}_{\text{no clust}}(\X^{C})$ is the BIC of the LCA model on the current clustering variables and $\text{BIC}(X^{P})$ is the BIC for a Multinomial distribution fit. Then $X^P$ is considered useful for clustering if the difference $BIC_A - BIC_B$ is greater than zero. To search through the model space, the authors propose a forward headlong search algorithm \citep{badsberg:1992} consisting of inclusion and removal steps. The algorithm has the advantage of being more computationally efficient than a backward search, but the disadvantage of being sensitive to the initialization of the set of clustering variables.

To deal with the problems of multimodality of the LCA log-likelihood and the sensitivity to the initialization, \cite{bartolucci:etal:2016} propose to add an extra step in the variable selection procedure of \cite{dean:raftery:2010}. They consider a \emph{random check} step aimed at initializing the estimation algorithm with a large number of random starting values, so as to prevent the problem of incurring in local optima. In an application to nursing home evaluation, the authors also perform a sensitivity analysis of the selected variables with respect to the initialization of the set of the clustering variables. The extra step is shown to be beneficial in increasing the chances of finding the global optimum.

\begin{figure}[!tb]
   \centering
    \includegraphics[scale=1]{graph_5.pdf}\quad\quad\includegraphics[scale=1]{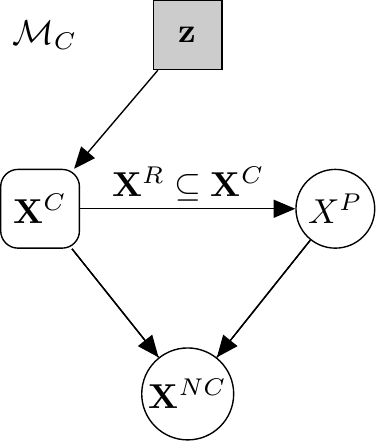}
   \caption{\label{graph5} The models proposed in \cite{fop:2017}.}
\end{figure} 

A major drawback of \cite{dean:raftery:2010} framework is the independence assumption between the proposed variable and the set of clustering ones in model $\mathcal{M}_B$. In fact, because of this assumption, the model does not take into account that the proposed variable could be redundant for clustering given the set of already selected relevant variables. This results in the method not being capable of discarding variables related to the clustering ones, but not directly related to the group structure itself. To overcome the problem, \cite{fop:2017} propose the modeling framework depicted in Fig.~\ref{graph5}. Following \cite{maugis:celeux:2009:a,maugis:celeux:2009:b} (consult Section~\ref{vsgmm_modsel}), in model $\mathcal{M}_C$ the proposed variable is not relevant for clustering and it is assumed to be dependent on the clustering variables $\X^C$. Model $\mathcal{M}_A$ is as before, while under model $\mathcal{M}_C$ the density of $\X$ is given by:
$$
\mathcal{M}_C\colon  \quad p(\X \vbar \z) = p(\X^{C} \vbar \z)\,p(X^P \vbar \X^R \subseteq \X^C) \, p(X^{NC}\vbar \X^C, \X^P),
$$
where the conditional distribution $p(X^P \vbar \X^R \subseteq \X^C)$ is modeled using a multinomial logistic regression. Only a subset $\X^R \subseteq \X^C$ of relevant predictors enter in the regression, avoiding the inclusion of extra parameters without necessarily increasing the likelihood. Moreover the subset $\X^R$ is allowed to be the empty set, and in the case of $\X^R=\varnothing$ the \cite{dean:raftery:2010} method is automatically recovered. For $\mathcal{M}_C$ the BIC is given by:
$$
\text{BIC}_C = \text{BIC}_{\text{no clust}}(\X^{C}) + \text{BIC}_{\text{reg}}(X^P \vbar \X^R \subseteq \X^C),
$$ 
with $\text{BIC}_{\text{reg}}(X^P \vbar \X^R \subseteq \X^C)$ the BIC of the multinomial logistic regression after the selection of the set $\X^R$, accomplished using a simple backward stepwise search, similarly to \cite{maugis:celeux:2009:a,maugis:celeux:2009:b}. Models are compared by means of the difference $\text{BIC}_A - \text{BIC}_C$ and $X^P$ is added to $\X^C$ if this quantity is greater than zero. To perform the model search, the authors suggest a backward \emph{swap-stepwise} selection algorithm where standard removal and inclusion steps are alternated to a \emph{swap step}. In this step two different configurations of model $\mathcal{M}_C$ are compared and they differ in the fact that one clustering variable is swapped with one of the non-clustering variables. The extra step allows the algorithm to not incur in local optima in the case of highly correlated variables. An application to the clustering of patients suffering low back pain shows that the method is capable of performing a parsimonious variable selection with a good classification performance and interpretable results. As in the case of GMMs, both \cite{dean:raftery:2010} and \cite{fop:2017} methods return the best clustering variables and the optimal number of latent classes; see the cited works for details.

\cite{toussile:2009} present an approach where the modeling problem of simultaneously selecting the number of latent classes and the relevant variables is considered in a density estimation framework. Let $\mathcal{S}$ be the set of clustering variables and let $\mathcal{M}_{(G, \mathcal{S})}$ denote a model of probability distributions such that:
$$
\mathcal{M}_{(G, \mathcal{S})}\colon \quad p(\X) = p(\X^{C})\, p(\X^{NC}),
$$
with $p(\X^{C})$ denoting a LCA model on the relevant variables and $p(\X^{NC})$ the Multinomial distribution model for the non-clustering variables. The aim is to select the couple $(G, \mathcal{S})$, which defines the model that generated the data. The selection procedure is implemented using an algorithm that first finds the optimal subset $\mathcal{S}$ for a range of possible values of $G$ and then selects the optimal $G$ for a fixed set of clustering variables. The authors employ penalized log-likelihood criteria for model selection and consistency results for BIC type criteria are derived. A further generalization and discussion is found in \cite{bontemps:toussile:2013}, where they derive a criterion that minimizes a risk function based on the Kullback-Leibler divergence of the estimated density with respect to the true
density. The general form of this penalized log-likelihood model selection criterion is given by:
$$
\ell_{(G, \mathcal{S})}^* - \kappa \dfrac{\mathcal{D}_{(G, \mathcal{S})}}{N},
$$
with $\ell_{(G, \mathcal{S})}^*$ the maximized log-likelihood under model $\mathcal{M}{(G, \mathcal{S})}$, $\mathcal{D}_{(G, \mathcal{S})}$ indicating the dimension of the model and $\kappa$ is a parameter depending on the sample size and the collection of the models. The authors suggest the use of \emph{slope heuristics} \citep[see][for an overview]{baudry:2012} to calibrate the value of $\kappa$.

\cite{marbac:sedki:arxiv} extend the approach of \cite{marbac:sedki:2017} (see Section \ref{vsgmm_modsel}) to clustering and variable selection of data of mixed type, with data on only categorical variables as a particular case. The method is based again on the MICL criterion and does not require multiple calls of the EM algorithm. In addition, the approach can manage situations where data have missing values. However, global and local independence need to be assumed to compute the MICL, and optimization is carried out over the variable indicator $\bm{\omega}$ and the cluster membership indicator $\mathbf{z}$.

\subsection{Other approaches}
As for GMMs, other methods have been suggested for variable selection in LCA. \cite{zhang:ip:2014} present a filter method where the absolute and relative importance of a variable towards the clustering structure is assessed by means of two measures. One is the \emph{expected posterior gradient}, which is a measure of the relative change between the entropy of the prior distribution of the latent classes and the entropy of the posterior distribution of the latent classes after observing the data. This quantity is bounded between 0 and 2 and higher values indicate higher discriminative power of a variable. The other measure is the \emph{Kolmogorov variation of posterior distribution}, which is based on the Kolmogorov distance between the class-conditional distributions. The measure is linked to the classification accuracy and the total variation of the posterior distribution, and, if there is a substantial reduction in this quantity without variable $X_j$ in the model, then $X_j$ can be considered a relevant clustering variable. Both measures are presented in the context of LCA for mixed type data, and the observation of only categorical data is a particular case. 

Another filter method where the selection is carried out by evaluating the impact of a variable on the clustering solution is proposed in \cite{bartolucci:2017}. Here, the initial set of variables is assumed to provide an optimal clustering of the data and the objective is to select a minimum subset of variables that is sufficient to reproduce the inferred classification. To achieve this, the authors suggest a selection algorithm that removes the variables whose presence does not significantly impact the clustering. Suppose a LCA model has been fitted to the data and the estimated posterior probabilities $\hat{u}_{ig}$ (consult Section~\ref{mbc}) have been obtained. In the procedure, for each variable $j$, the proportion $\mathcal{F}_{-j}$ of observations whose class assignment is modified when $X_j$ is removed with respect to the initial clustering is computed. By~\eqref{lca}, the updated estimate $\hat{u}^{(-j)}_{ig}$ after removing $X_j$ is simply given by:
$$
\hat{u}^{(-j)}_{ig} = \dfrac{\hat{\tau}_g\prod_{\substack{h=1\\h\neq j}}^J \prod_{c=1}^{C_h} \hat{\pi}_{ghc}^{\mathds{1}\lbrace x_{ih} = c \rbrace}}{\sumg \hat{\tau}_g\prod_{\substack{l=1\\l\neq j}}^J \prod_{c=1}^{C_l} \hat{\pi}_{glc}^{\mathds{1}\lbrace x_{il} = c \rbrace}}.
$$
Therefore:
$$
\mathcal{F}_{-j} = \dfrac{1}{N} \sumi \mathds{1}\lbrace \text{MAP}(\hat{\mathbf{u}}^{(-j)}_{i}) \neq \text{MAP}(\hat{\mathbf{u}}_{i})  \rbrace
$$
Then the variable with the minimum $\mathcal{F}_{-j}$ is removed. If some variables have the same value of $\mathcal{F}_{-j}$, the variable to be removed is the one with the minimum Kullback-Leibler distance:
$$
\sumi \sumg \hat{u}_{ig} \log \dfrac{\hat{u}_{ig}}{\hat{u}^{(-j)}_{ig}}.
$$
The method is developed in connection to item selection for questionnaires and is fast and simple to implement. However, it requires the somewhat strong assumption that the inferred classification and the number of latent classes does not change when removing the variables.

It is worth to mention an approach that lies outside the modeling framework of LCA and is closely related to the work of \cite{hoff:2006}. \cite{hoff:2005} presents a method for subset clustering of binary sequences where each latent class is characterized by its own set of discriminating variables that differentiates it from the rest of the classes. Let $\varphi_{gj}$ be a binary variable indicating the relevance of variable $j$ to cluster $g$. The author suggests to parameterize the class conditional probability of occurrence of $X_j$ as:
$$
\pi_{gj} = \varphi_{gj} \tilde{\pi}_{gj} + (1 - \varphi_{gj}) \pi_j,
$$
where $\tilde{\pi}_{gj}$ is the probability that $X_j=1$ within class $g$ and $\pi_j$ is the probability of observing variable $X_j$ in the data. Thus, if $\varphi_{gj}$ is active, the corresponding variable is relevant and the $\pi_{gj}$ differs from the overall data value. The framework involves a P\'olya urn scheme for the parameters and the indicator variables and results in a Dirichlet process mixture model. Inference is performed using MCMC and allows to recover the subset of informative variables of each latent class.

\subsection{R packages and data example}

The \texttt{R} packages for variable selection for latent class analysis are \texttt{ClustMMDD} \citep{clustmmdd}, \texttt{LCAvarsel} \citep{lcavarsel} and \texttt{VarSelLCM} \citep{varsellcm}. In particular, we note that \texttt{VarSelLCM} implements a more general framework for clustering and variable selection of data of mixed type \citep{marbac:sedki:arxiv}. Table~\ref{pack_LCA} lists the packages, with information regarding the type and the method; all implement a model selection approach.
  
\begin{table}[!b]
\centering
\caption{\label{pack_LCA} \textsf{\em R} packages for LCA variable selection.}
 \begin{tabular}{lll}
 \toprule
  \bf Package	&	\bf Type	&	\bf Method\\
  \midrule
  \multirow{2}{*}{\texttt{ClustMMDD}}	&	\multirow{2}{*}{model selection}	&	\cite{toussile:2009}\\
					&						&	\cite{bontemps:toussile:2013}\\[1em]
  \multirow{2}{*}{\texttt{LCAvarsel}}	&	\multirow{2}{*}{model selection}	&	\cite{dean:raftery:2010}\\
					&						&	\cite{fop:2017}\\[1em]
  \multirow{2}{*}{\texttt{VarSelLCM}}	&	\multirow{2}{*}{model selection}	&	\cite{marbac:sedki:2017}\\
					&						&	\cite{marbac:sedki:arxiv}\\
  \bottomrule
 \end{tabular}
\end{table}

\begin{figure}[bt]
\centering
  \includegraphics[scale=0.62]{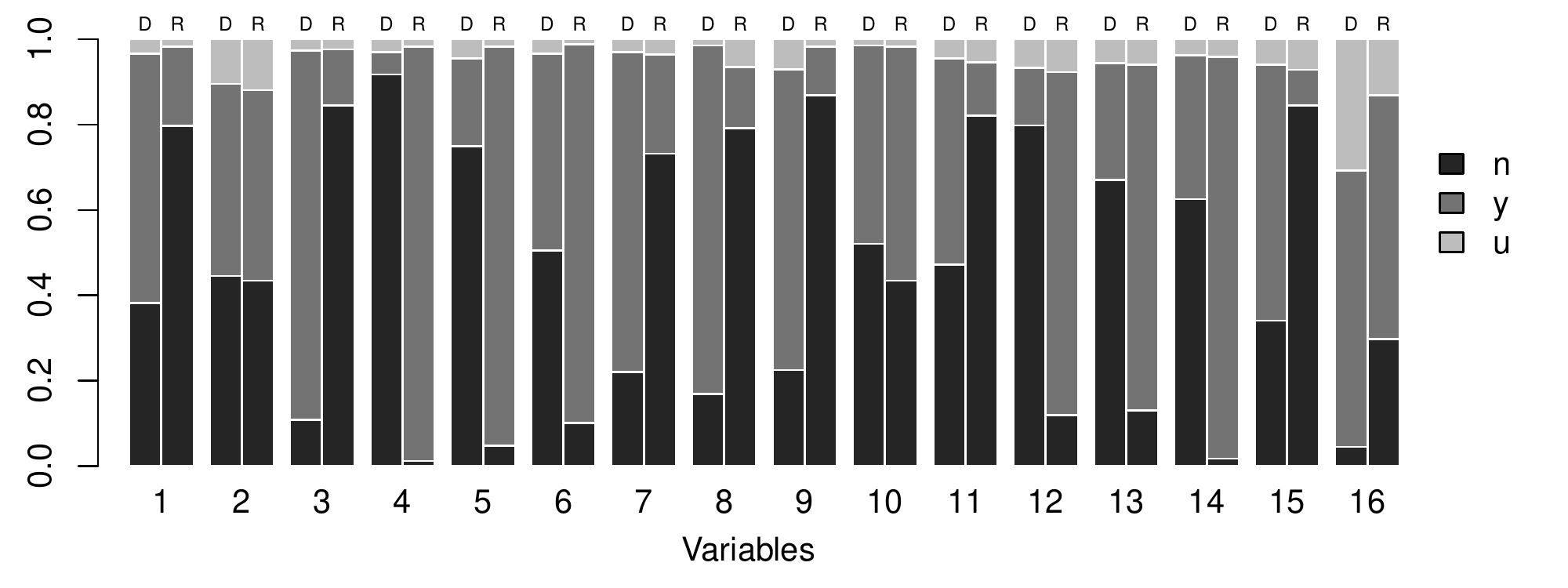}
  \caption{\label{fig_uscongress} Key votes in the {\em\texttt{uscongress}} data. For each vote, a paired stacked-barplot represents the observed relative frequency of a response \emph{(\texttt{n})}, \emph{(\texttt{y})} or \emph{(\texttt{u})} within the Democrats \emph{(\textsf{D})} and the Republicans \emph{(\textsf{R})} parties.}
\end{figure}

In this section we consider the \texttt{uscongress} data available in the UCI repository at the web page \url{https://archive.ics.uci.edu/ml/datasets/congressional+voting+records}. The data contain votes of $N = 435$  members of the U.S. House of Representatives, 267 Democrats and 168 Republicans. Each record expresses the position of a member on 16 key votes regarding major issues selected by the \cite{uscongress}; the votes are presented in Table~\ref{us}. A vote can take three possible outcomes: \texttt{y} (yea), if in favor of a specific matter, \texttt{n} (nay), if against, or \texttt{u} (unknown), if the position is unknown. A graphical representation of the data is in Figure~\ref{fig_uscongress}. The observations are classified into two groups corresponding to the main parties, but it is reasonable to expect the presence of more than two classes in the data because of internal subdivisions. Indeed, at the time the Democratic party was split into Northern and Southern Democrats, and a large portion of the second often voted in agreement with the Republican party \citep{uscongress}. 

\begin{table}[tb]
\centering
\caption{\label{us} Variables in the \emph{\texttt{uscongress}} dataset.}
\begin{tabular}{ll}
\toprule
1. Handicapped infants	&	9. MX missile\\
2. Water project cost-sharing	&	10. Immigration\\	
3. Adoption of the budget resolution	&	11. Synfuels corporation cutback\\
4. Physician fee freeze	&	12. Education spending\\
5. El Salvador aid	&	13. Superfund right to sue\\
6. Religious groups in schools	&	14. Crime\\
7. Anti-satellite test ban	&	15. Duty-free exports\\
8. Aid to Nicaraguan contras	&	16. Export administration act/South Africa\\
\bottomrule
\end{tabular}
\end{table}

First, we fit a LCA model on all the variables, determining the number of components using BIC. The selected model is a 4-class model and Table~\ref{cross-all} contains a cross-tabulation of the estimated classification and the party membership. Class 1 and Class 2 are fairly polarized into Democrats and Republicans. Class 3 is predominantly characterized by Democrats, while Class 4 contains an equal amount of representatives from both sides.

\begin{table}[!b]
\parbox{.45\linewidth}{
\centering
\caption{\label{cross-all} Cross-tabulation between the classification estimated by the LCA model on all the variables and the political affiliation.}
\begin{tabular}{lcccc}
& \multicolumn{4}{c}{\bf Class}\\
 & 1 & 2 & 3 & 4 \\ 
 \midrule
  Dem. & 170 &  23 &  65 &  9 \\ 
  Rep. &   1 & 132 &  25 &  10 \\ 
 \bottomrule
\end{tabular}
}
\hfill
\parbox{.45\linewidth}{
\centering
\caption{\label{cross-selswap} Cross-tabulation between the classification inferred on the variables selected by \texttt{\em LCAvarsel} procedure and the political affiliation.}
\begin{tabular}{lcccc}
& \multicolumn{4}{c}{\bf Class}\\
 & 1 & 2 & 3 & 4 \\ 
 \midrule
  Dem. & 159 &  29 &  74 &   5 \\ 
  Rep. &   3 & 129 &  30 &   6 \\ 
 \bottomrule
\end{tabular}
}
\end{table}

\begin{figure}[t]
\centering
  \includegraphics[scale=0.6]{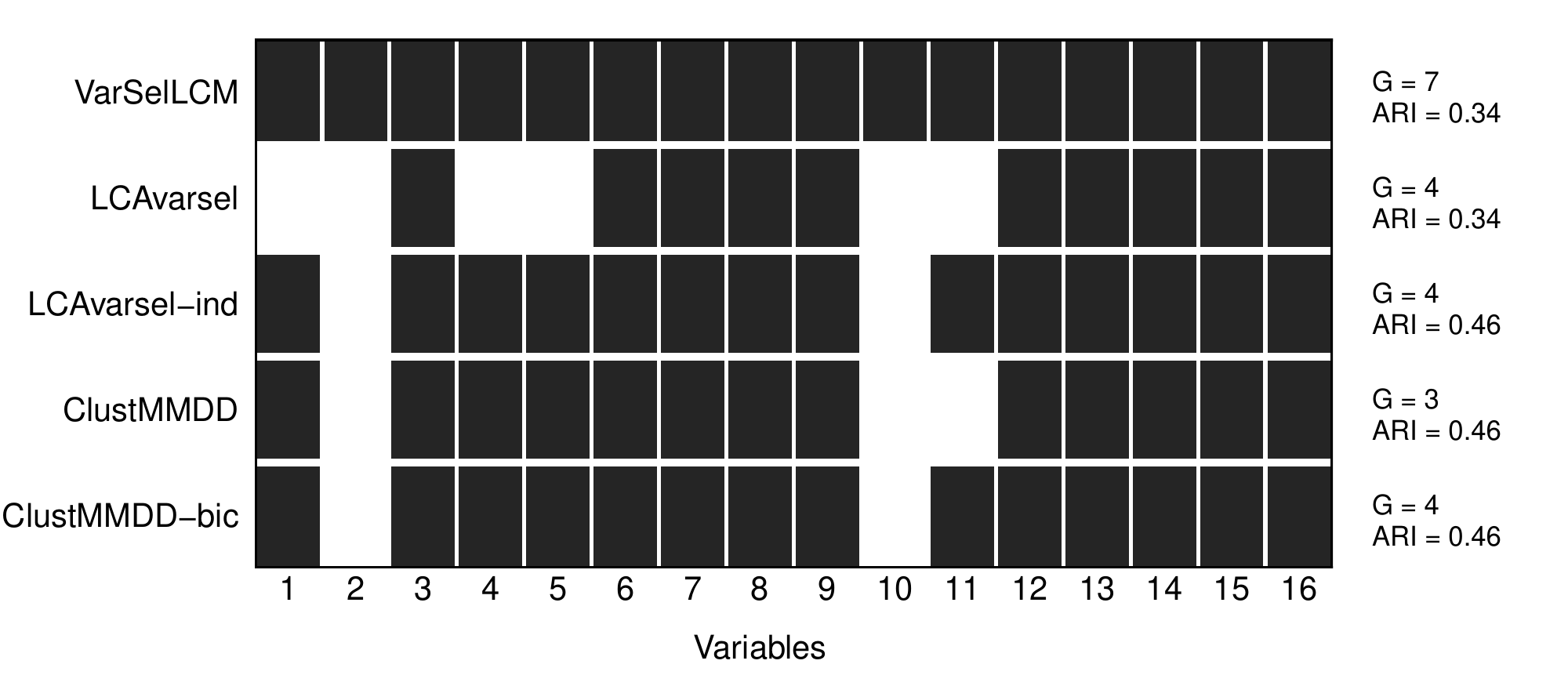}
  \caption{\label{fig_lca_varsel_1} Variables selected via the \texttt{\em ClustMMDD-bic}, \texttt{\em ClustMMDD}, \texttt{\em LCAvarsel-ind}, \texttt{\em LCAvarsel} and \texttt{VarSelLCM} methods. A dark square indicates the variable has been selected. For each method, also the selected number of latent classes and the ARI between the estimated classification and the party affiliation are reported.}
\end{figure}

\begin{figure}[!h]
\centering
    \subfloat[][\texttt{ClustMMDD-bic}]
    {\includegraphics[width=.45\textwidth]{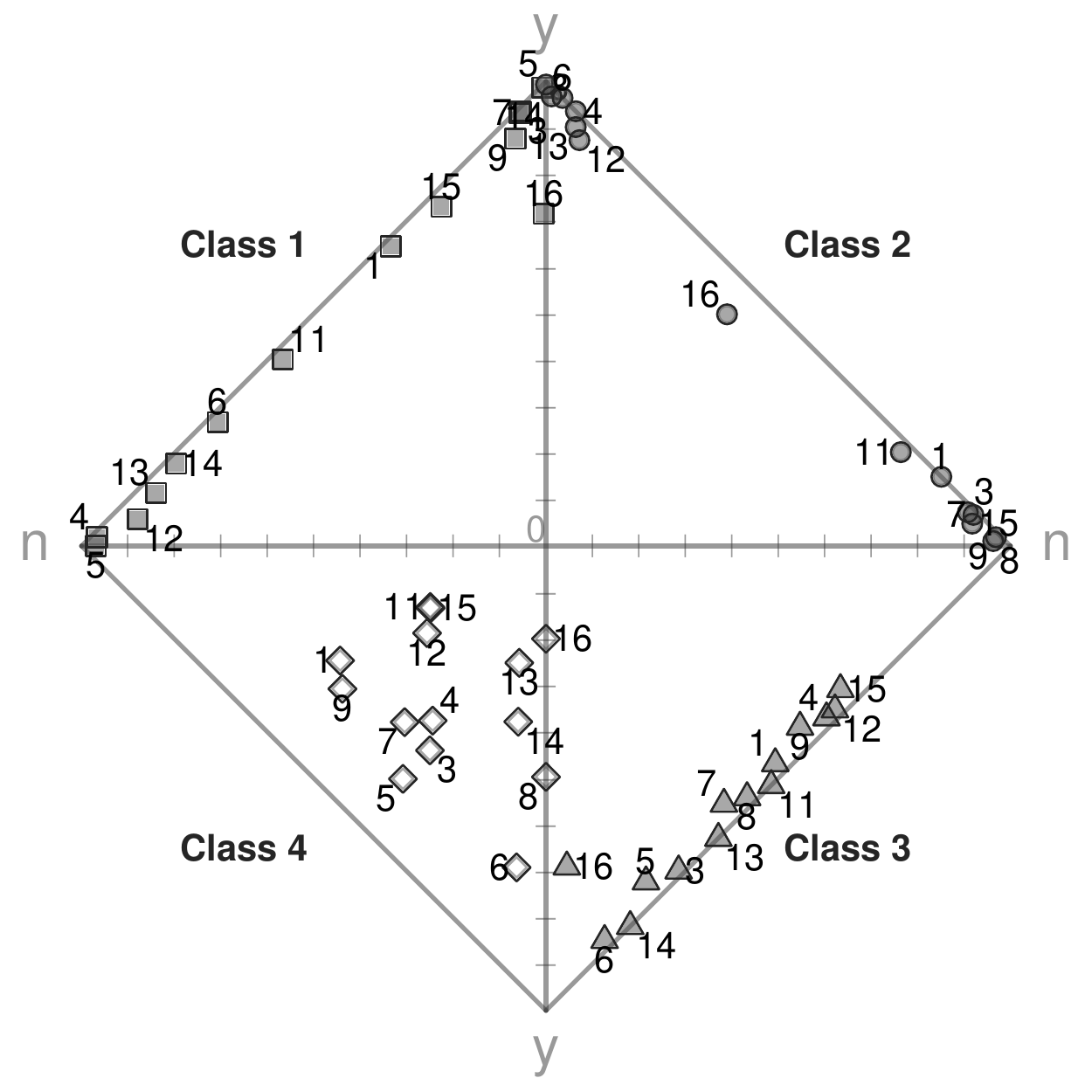}} \quad
    \subfloat[][\texttt{ClustMMDD}]
    {\includegraphics[width=.45\textwidth]{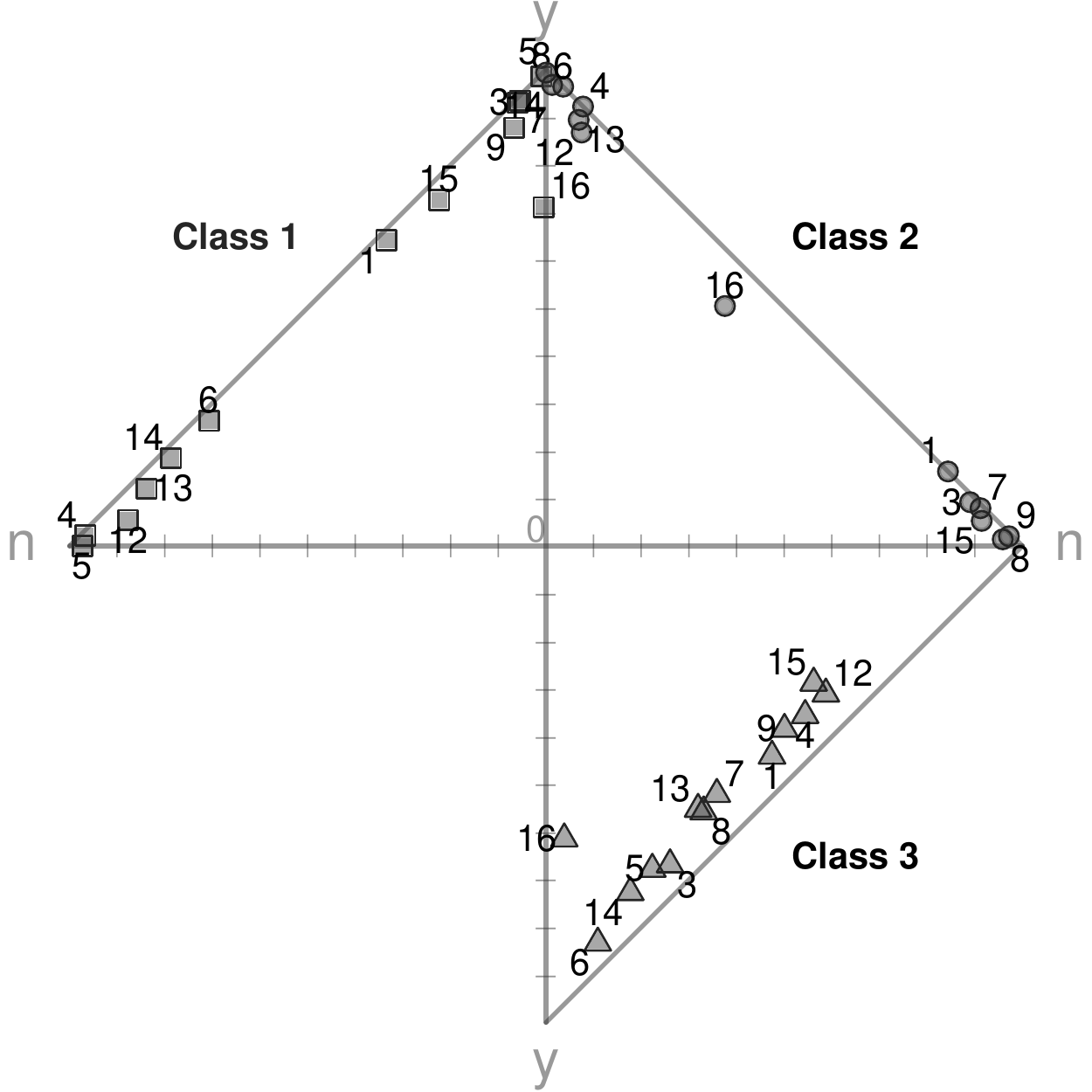}} \\
    \subfloat[][\texttt{LCAvarsel-ind}]
    {\includegraphics[width=.45\textwidth]{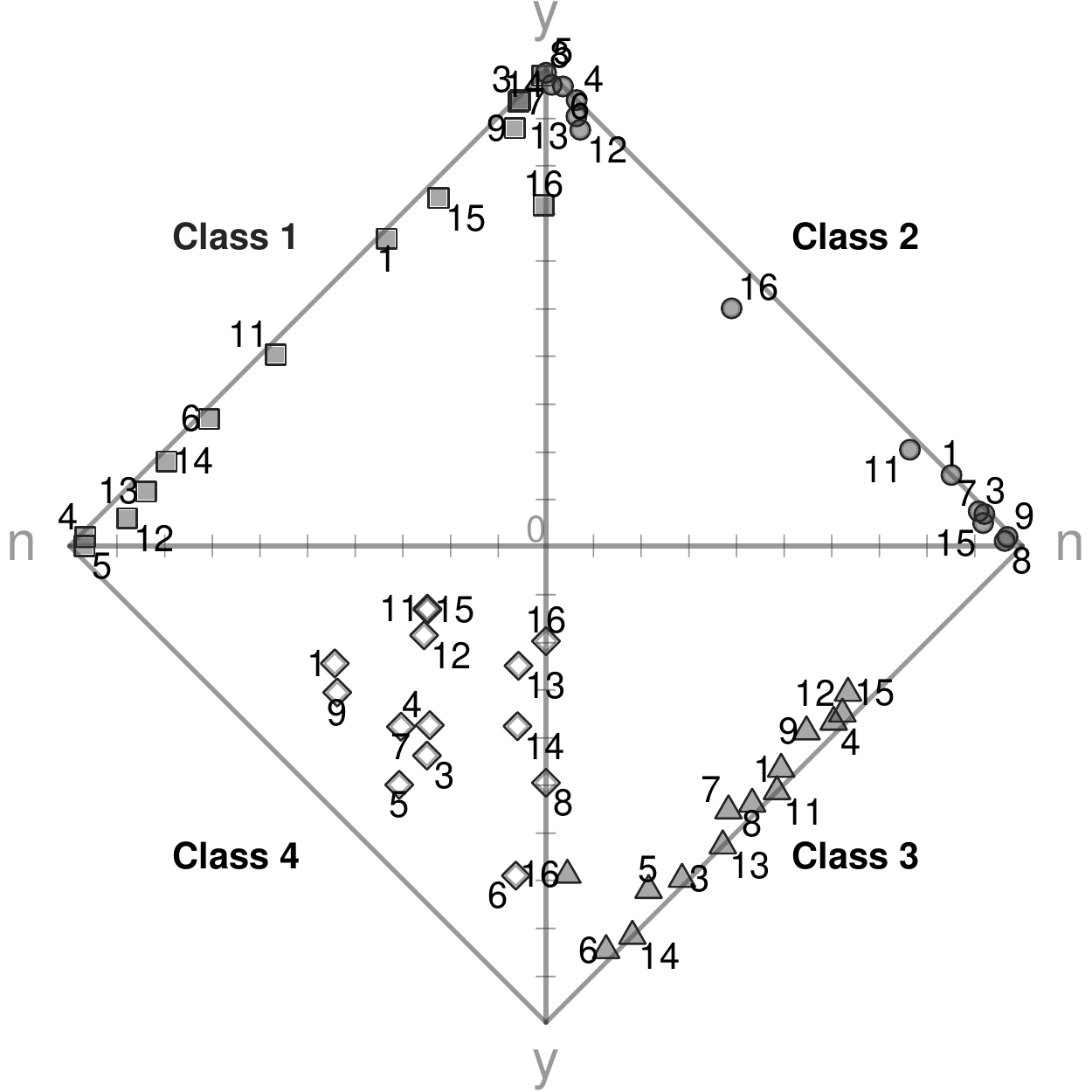}} \quad
    \subfloat[][\texttt{LCAvarsel}]
    {\includegraphics[width=.45\textwidth]{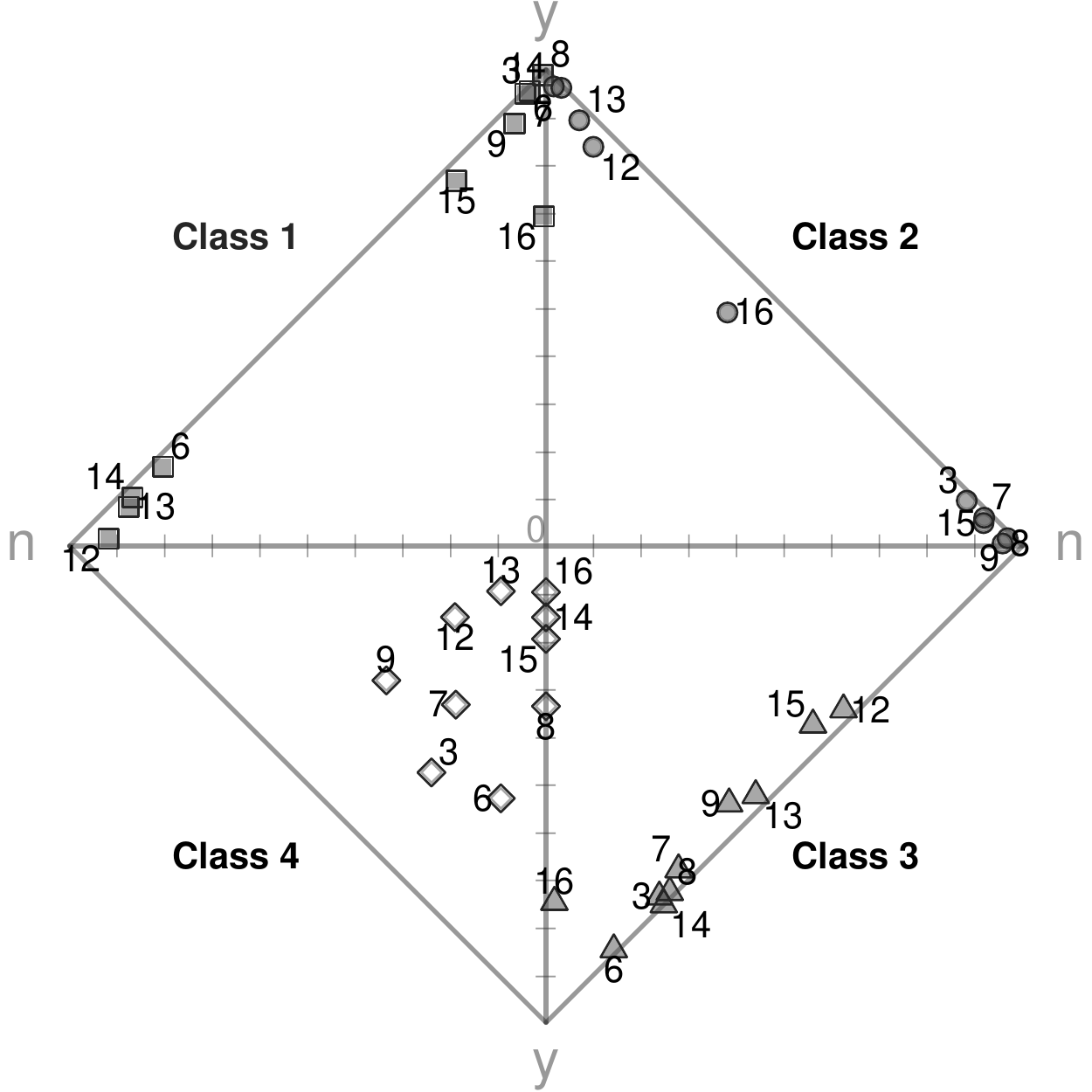}}
    \caption{\label{fig_lca_varsel_2} Class-conditional probabilities of the outcomes for the votes selected by \texttt{\em ClustMMDD-bic}, \texttt{\em ClustMMDD}, \texttt{\em LCAvarsel-ind} and \texttt{\em LCAvarsel}. Each quadrant contains the 2-dimensional simplex representing, for every selected variable, the class-conditional probabilities of outcome \texttt{\em n}, \texttt{\em y} and \texttt{\em u} by difference. All the quadrants have values ranging from 0 to 1.  Points close to the axis vertexes denote high probability of outcome \texttt{\em n} or \texttt{\em y}, while points close to the center indicate high probability of outcome \texttt{\em u}. }
\end{figure}

Then we apply the variable selection procedures \texttt{ClustMMDD-bic} \citep{toussile:2009}, \texttt{ClustMMDD} \citep{bontemps:toussile:2013}, \texttt{LCAvarsel-ind} \citep{dean:raftery:2010}, \texttt{LCAvarsel} \citep{fop:2017} and \texttt{VarSelLCM} \citep{marbac:sedki:2017,marbac:sedki:arxiv}. The variables selected by each method are displayed in Figure~\ref{fig_lca_varsel_1}. Also the chosen number of latent classes and the ARI between the estimated classification and the party affiliation are reported. \texttt{VarSelLCM} does not discard any variable in this example. The method underlying the package makes use of both local and global independence assumptions, and they are too restrictive for this type of data. \texttt{ClustMMDD} and \texttt{LCAvarsel-ind} select the same model and discard 2 variables. \texttt{ClustMMDD} discards 3 variables and selects a 3-component model. Note that the common discarded variables 2 and 10 are votes with close call results, and thus likely uninformative with respect to the two parties or any clustering structure. \texttt{LCAvarsel} performs a more parsimonious selection, choosing a model with 4 latent classes and retaining 10 variables, but attaining a lower ARI.  Figure~\ref{fig_lca_varsel_2} displays the class-conditional probabilities of the outcomes for the votes selected by \texttt{ClustMMDD-bic}, \texttt{ClustMMDD}, \texttt{LCAvarsel-ind} and \texttt{LCAvarsel}; we did not include \texttt{VarSelLCM}, as it discarded none of the variables. 
We focus the attention on the \texttt{LCAvarsel} result, as it is the package performing the most parsimonious selection. Table~\ref{cross-selswap} reports a tabulation of the estimated classification and the party affiliation, with an interpretation similar to Table~\ref{cross-all}. Figure~\ref{fig_lca_varsel_2} (d) suggests that Class 1 and 2 are denoted by very polarized outcomes and opposite voting positions. Class 3 includes members who likely expressed a vote not loyal to their party line. Class 4 is characterized by an higher probability of an unknown position regarding the selected key votes. 

Table~\ref{time-lca} reports the computing time of the packages listed in Table~\ref{pack_LCA} on a Dell machine with Intel Core i7-3770 CPU @3.40GHz$\times$8. The methods are listed from the fastest to the slowest. \texttt{LCAvarsel-ind} is the fastest, followed by \texttt{VarSelLCM}, although the last did not discard any of the variables. \texttt{LCAvarsel} is slower, but still with an acceptable computing time compared to the two. Methods \texttt{ClustMMDD} and \texttt{ClustMMDD-bic} are the slowest in this example and have the same computing time, since they implement the same selection procedure. The algorithm at the basis of package \texttt{ClustMMDD} performs a more extensive search than the greedy ones implemented in \texttt{LCAvarsel} and \texttt{VarSelLCM}, however at a larger computational cost. Lastly, it is worth to notice that packages \texttt{LCAvarsel} and \texttt{VarSelLCM} can implement parallel computations.

\begin{table}[!bt]
\centering
\caption{\label{time-lca} Computing time (in seconds) and relative computing time of the R packages for variable selection for LCA.}
\begin{tabular}{lrr}
\toprule
\bf Package	&	\bf Time (sec.)	&	\bf Relative \\
\midrule
\texttt{LCAvarsel-ind} 	& 114.94 & 1.00 \\ 
\texttt{VarSelLCM} 	& 211.20 & 1.84 \\ 
\texttt{LCAvarsel}	& 591.62 & 5.15 \\ 
\texttt{ClustMMDD}  & 11585.65 & 100.80 \\ 
\texttt{ClustMMDD-bic}  & 11585.65 & 100.80 \\ 
\bottomrule
\end{tabular}
\end{table}

\section{Discussion}
\label{disc}

Stimulated by the wide popularity of the approach and the diffusion of high-dimensional data, the topic of variable selection for model-based clustering has seen increasing attention and rapid development. In this paper we gave an overview of the available variable selection methods, starting from early works to the most recent state of the art. We suggested a general and systematic picture of the features characterizing the different methods. The exposition focused on variable selection methods for Gaussian mixture models and latent class analysis, the most common model-based clustering approaches. Illustrative data examples have been used to show some of the methods in action and we provided references to the available \textsf{R} packages implementing them. The datasets and the R code used for the analyses is available at the web page \url{https://github.com/michaelfop}. 
We conclude with some final remarks and comments.

\subsubsection*{Distributional assumptions} 
Assumptions concerning the dependence among relevant and irrelevant variables play a crucial role in the variable selection procedure. The global independence assumption simplifies the model for the joint distribution of irrelevant and relevant variables. Directly or indirectly, Bayesian and penalization approaches make use of this assumption, resulting in frameworks with a simpler association structure and that allow to discard uninformative variables. However, in the case where the variables are highly correlated, the global independence assumption can prevent to accomplish a parsimonious selection, in which also redundant variables are discarded \citep{law:figueiredo:2004,tadesse:sha:2005,raftery:dean:2006}. On the other hand, model selection methods allow to depict a flexible and realistic structure for the relations among relevant, redundant and uninformative variables, but at the cost of a more complex model for the joint distribution of relevant and irrelevant variables. 
We remark that, when using a variable selection method, the trade-off between model complexity and selection performance needs to be taken into account.

The local independence assumption is commonly used in latent variable models, especially in the case of multivariate categorical data clustering. This assumption notably simplifies the model for the joint distribution of the relevant clustering variables \citep{bartholomew:knott:2011}. However, it can hardly hold in practice and several approaches in the literature have been proposed to overcome it. For example, \cite{gollini:murphy:2014} propose a setting with continuous latent variables that allow to model the dependences among the observed categorical variables; \cite{marbac:biernacki:2015} develop a framework where the categorical variables are grouped into blocks, each one following a specific distribution that takes into account the dependency between variables. Extending these frameworks to allow for variable selection would be interest of future research.

In this review, we mainly focused on Gaussian and Multinomial mixture models for clustering of continuous and categorical data. Variable selection for model-based clustering with other distributions and/or of data of different nature is still an open research area. The topic has only recently started to attract attention: \cite{wallace:2017} propose a framework for clustering and variable selection with the multivariate skew-Normal distribution, while \cite{marbac:sedki:arxiv} define a method for data of mixed types. There is certainly a wide scope for further investigations and developments in this direction.

\subsubsection*{Computational aspects}
Solving a variable selection problem is a task that requires a noticeable computational cost. For fixed number of mixture components, in general there are $2^J$ possible combinations of variables that could be considered as relevant clustering variables \citep{miller:2002}. The problem is worsened when the variable selection is concomitant to the problem of mixture components selection and model estimation. As already noted, filter approaches keep the two tasks separated, since the variable selection procedure is executed before or after estimation of the clustering model. Wrapper approaches combine model estimation, variable selection and number of components determination; hence, usually multiple models need to be estimated for different combinations of clustering variables and number of components. For these reasons, the firsts are usually faster, while the seconds are computationally expensive, but often give superior results \citep{guyon:2003}. To speed up the computations, a filter method could be employed as a preliminary step to reduce the number of variables, then the wrapper method could be used on the reduced set of variables. 

Regarding the statistical approach used to perform the variable selection, the various methods present different computational characteristics and issues. Bayesian methods provide a solid ground for uncertainty evaluation of the variable selection process. However, the MCMC schemes employed within this class of methods  often require runs with a large number of iterations to explore the enormous space and ensure convergence \citep[][indeed, all the works examined in this review considered a number of iterations in the order of $10^5-10^6$]{tadesse:sha:2005}. Unfortunately, the MCMC algorithms used do not allow parallelization of the computations in order to speed up the process. Model selection approaches usually implement greedy stepwise algorithms that have a general computational complexity of $\mathcal{O}(KJ^2)$, hence they become rapidly impractical as the number of variables and mixture components increases. If the number of variable is large, and is expected that only a small number of variables $J_0 \ll J$ are relevant, forward-type algorithms can be applied, thus reducing the complexity to $\mathcal{O}(KJJ_0)$. However, in this case the difficult issue of how to initialize the algorithm and the set of clustering variables arises. Backward-type procedures do not require initialization of the clustering set, but can be very computationally demanding if the actual set of clustering variables is small. Headlong strategies could be employed to mitigate the problem \citep{badsberg:1992}. Nonetheless, the advantage of these stepwise methods is that they can be implemented in parallel, saving computational time. Penalization approaches are usually faster than Bayesian and model selection methods. In fact, the computational complexity of these approaches mainly depends on the form of the penalty function adopted and, in most cases, analytical solution to the optimization problem or efficient numerical procedures are already available. For this reason, they tend to scale particularly well to high-dimensional settings and have been proven to perform well also in the case of data with thousands of variables. Nevertheless, these methods do not allow for parallel computations.

Given the constant increase of the data dimensionality, the development of efficient algorithms allowing for scalable model-based clustering and variable selection is a relevant research topic that is likely to attract considerable attention in the future.

\subsubsection*{Missing data}
Missing observations are a common issue in many data analysis problems. Among the works examined in this review, only few present methods for model-based clustering and variable selection in the presence of missing data. \cite{maugis:2012} extend the model selection method of \cite{maugis:celeux:2009:b} with the aim of avoiding a preliminary imputation procedure. The authors consider the case of missing values generated under the \emph{missing at random} mechanism \citep[][MAR]{little:2002}. Under this assumption, the missing responses are ignorable for likelihood-based inference. Hence, the framework and the model selection criterion can be stated in terms of only the observed data and the imputation process is avoided. \cite{bartolucci:etal:2016} use similar arguments for clustering categorical data via the latent class analysis model. The same MAR assumption is also considered in \cite{marbac:sedki:arxiv}, where the MICL criterion used for variable selection is computed considering only the observed entries. In application to data related to the quality-of-life of elderly patients hosted in nursing homes, \cite{bartolucci:2017} move away from the MAR assumption and suggest to add an extra category corresponding to the missing responses, circumventing the assumption of ignorability of the missing data.

In general, in the literature multiple works exist for mixture model estimation with missing data, see for example: \cite{mclachlan:peel:2000,little:2002,hunt:2003, formann:2007,chen:2014}. These approaches could be incorporated in a general framework for model-based clustering and variable selection for data with missing entries and may be interest of future developments.

\subsubsection*{Software}
The last remark regards software availability. Despite the quantity of theory and methods developed, not many \textsf{R} packages for variable selection in model-based clustering are available. In particular, there is lack of packages implementing the various penalization methods for GMMs, especially useful for clustering high-dimensional data. This is somewhat surprising given the practical importance of the topic.\\

{\bf Acknowledgments.} We would like to thank the Editor, Associate Editor and Referees whose suggestions and comments helped to improve the quality of the work.

\bibliographystyle{imsart-nameyear}
\bibliography{bibliography.bib}

\end{document}